\documentclass[useAMS,usenatbib]{mn2e}

\usepackage{graphicx}
\usepackage{amsmath}
\usepackage{txfonts}
\usepackage{textcomp}
\usepackage{subfigure}
\usepackage{color}
\usepackage[normalem]{ ulem }
\usepackage{soul}
\usepackage{braket}

\title[]{Collisional excitation of doubly and triply deuterated
  ammonia ND$_2$H and ND$_3$ by H$_2$.}

\author[F. Daniel et al.]{F. Daniel$^{1}$\thanks{E-mail:
fabien.daniel@obs.ujf-grenoble.fr}, 
C. Rist,$^{1}$
A. Faure,$^{1}$ 
E. Roueff,$^{2}$
M. G\'erin,$^{3}$
D.C. Lis,$^{4,5}$
P. Hily--Blant,$^{1}$ 
\newauthor
A. Bacmann$^{1}$
and
L. Wiesenfeld$^{1}$
\\
$^{1}$ IPAG, Observatoire de Grenoble, Universit\'e Joseph Fourier, CNRS UMR5571, B.P. 53, 38041 Grenoble Cedex 09, France \\
$^{2}$ LERMA, Observatoire de Paris, PSL Research University, CNRS, Sorbonne Universit\'es, UPMC Univ. Paris 06, F-92190, Meudon, France \\
$^{3}$  LERMA, Observatoire de Paris, PSL Research University, CNRS, Sorbonne Universit\'es, UPMC Univ. Paris 06, Ecole normale sup\'erieure, F-75005, Paris, France \\
$^{4}$  LERMA, Observatoire de Paris, PSL Research University, CNRS, Sorbonne Universit\'es, UPMC Univ. Paris 06, F-75014, Paris, France  \\
$^{5}$ California Institute of Technology, Cahill Center for Astronomy and Astrophysics 301-17, Pasadena, CA 91125, USA }

\begin{document}

\date{Accepted XXX. Received XXX; in original form XXX}

\pagerange{\pageref{firstpage}--\pageref{lastpage}} \pubyear{2014}

\maketitle

\label{firstpage}

\begin{abstract}

The availability of collisional rate coefficients is a prerequisite
for an accurate interpretation of astrophysical observations, since the
observed media often harbour densities where molecules are
populated under non--LTE conditions. In the current study, we present
calculations of rate coefficients suitable to describe the various
spin isomers of multiply deuterated ammonia, namely the ND$_2$H and
ND$_3$ isotopologues. These calculations are based on the most
accurate NH$_3$--H$_2$ potential energy surface available,
which has been modified to describe the geometrical
changes induced by the nuclear substitutions. The dynamical
calculations are performed within the close--coupling formalism and
are carried out in order to provide rate coefficients up to a
temperature of $T$ = 50K. For the various isotopologues/symmetries, we
provide rate coefficients for the energy levels below $\sim$ 100
cm$^{-1}$. Subsequently, these new rate coefficients are used in
astrophysical models aimed at reproducing the NH$_2$D, ND$_2$H and
ND$_3$ observations previously reported towards the prestellar cores
B1b and 16293E. We thus update the estimates of the corresponding
column densities and find a reasonable agreement with the previous
models. In particular, the ortho--to--para ratios of NH$_2$D and NHD$_2$ are found
to be consistent with the statistical ratios.
\end{abstract}

\begin{keywords}
molecular data -- molecular processes -- scattering
\end{keywords}

\section{Introduction}

Ammonia is an ubiquitous molecule in space, which has been observed in
a large variety of media.  In some of these places, the temperature
can fall to a few kelvins, as in low mass star forming regions.  Others,
as for example shocked media,
have temperatures as high as a few thousand kelvins.
  Depending on the temperature, the chemistry that leads
to ammonia formation can follow different paths. Given the importance
of this molecule, in particular in the context of nitrogen chemistry,
much effort has been made in the past in order to characterize the
corresponding formation or destruction mechanisms. This proved to be
successful and nowadays, most of the observations can be, at least
qualitatively, understood. In some cases, however, astrochemical
models and observations still show discrepancies, as in the case of
circumstellar envelopes where the models under--predict the observed
molecular abundances by a few orders of magnitude
\citep{menten2010}. Additionally, in recent years, it was
found that the nuclear--spin states of a given molecule are not
necessarily populated according to simple statistical
considerations. This is the case for ammonia, for which some
observations report ortho--to--para ratios (OPR) in the range 0.5--0.7
\citep{hermsen1988,persson2012}, while a statistical ratio of 1 was
expected in the past, for a formation in the gas phase. If ammonia is formed in
the solid phase, and subsequently released from the grains, this ratio
would be $\sim$2. Recently, astrochemical gas-phase models  
have been
updated by including rigorous nuclear-spin selection rules relevant to
describe the ammonia formation. It was found that ratios lower than 1
can be well accounted for by such models in the low temperature regime,
where the H$_2$ gas is para-enriched \citep{rist2013,faure2013,sipila2015}.

While NH$_3$ is observed in media that span a  wide range of physical
conditions, the detection of its multi--deuterated counterparts,
i.e. ND$_2$H and ND$_3$, is restricted to cold regions. The
reasons behind this are now well established and lie in the fact that
the deuterium enrichment comes from the difference of zero point
energies associated with the hydrogenated or deuterated
isotopologues. These differences create a dichotomy in the chemical
network. Indeed, the reactions that lead to deuterated species have, in general,
exothermicities which are enhanced by comparison to the equivalent
reactions that involve their hydrogenated counterparts. This effect
creates a significant deuterium enrichment at low temperatures 
if H$_2$ is mainly in its para form. Indeed,
in the high temperature regime, the dichotomy between deuterated and
hydrogenated molecules vanishes, since the difference of
exothermicities become negligible with respect to the kinetic temperature.
In the particular case of ammonia, the main formation route consists
 of a sequence of reactions involving H$_2$, which is initiated by the  
 N$^+$ + H$_2$ $\to$ NH$^+$ + H reaction and ends with the dissociative recombination
 reaction NH$_4^+$ + $e^-$ $\to$ NH$_3$ + H \citep[see e.g.][]{dislaire2012}. 
 The initiating reaction has a small endothermicity \citep[in the range 100--300K, see e.g.][]{gerlich1993}. On the other
 hand, the endothermicity of the equivalent reaction involving HD and leading to ND$^+$ is reduced \citep[20--70K, see][]{marquette1988,sunderlin1994}, which
 hence favors the deuteration processes at low temperatures. 
Finally, the abundances of the NH$_3$ isotopologues decrease with the
number of deuterium atoms. Given the physical conditions
characteristics of low--mass star forming regions, which harbour low
gas temperatures (T $\lesssim$ 20K) and  relatively low gas densities 
(n(H$_2$) $\lesssim$ $10^7$ cm$^{-3}$), the low abundance implies that the
signal associated with these isotopologues is weak, which makes the
ND$_2$H and ND$_3$ molecules hard to detect with the current
facilities. Hence, so far, the detection of all four deuterated
istopologues of ammonia is restricted to two astrophysical objects,
the prestellar cores 16293E\footnote{This prestellar core is sometimes referred as L1689N in the literature} and Barnard 1 \citep{roueff2005}. 
This census has still to be enlarged to other objects to ensure statistically meaningful
conclusions, starting with objects where ND$_3$ has already been observed, as in NGC 1333 \citep{vandertak2002}.

The current work is aimed at providing collisional rate coefficients
for the ND$_2$H and ND$_3$ molecules, with p--H$_2$ as a collisional
partner. For these molecules, earlier calculations were performed with
He or H$_2$ with a reduced basis set
\citep{machin2006,machin2007,yang2008,wiesenfeld2011}.  Such rate coefficients are
essential in order to interpret the emission observed in
astrophysical objects and imply solving a problem of quantum
dynamics.  Recently, similar calculations \citep{ma2015,tkac2015} were
performed independently for NH$_3$ and ND$_3$ and using the Potential
Energy Surface (PES) reported by \citet{maret2009}, which is also used
in the current work.  However, rather than providing collisional rate
coefficients, the focus of \citet{ma2015} or \citet{tkac2015} was to
assess the possibility of testing the accuracy of the PES by comparing
theoretical predictions based on the PES with experimental data.  The
comparisons performed so far show a good agreement between experiments
and theory at collision energies above 430~cm$^{-1}$ \citep{tkac2015}.

The present paper is organized as follows. The potential energy surface is
described in Section \ref{potentiel} and the collisional dynamics
based on this surface in Section \ref{dynamique}.  We then describe
the rate coefficients in Section \ref{rates}. In Section
\ref{discussion}, we discuss the current results with respect to other
related collisional systems.  In Section \ref{astro}, we use our new
rate coefficients in order to re--interpret the observations
previously reported toward 16293E and Barnard 1.  Finally, we present
our conclusions in Section \ref{conclusion}.

\section[]{Potential Energy Surface (PES)} \label{potentiel}

The collisional excitation of an isotopic homologue takes place on the
same Born-Oppenheimer PES as the main isotopologue. All differences
are therefore contained in the dynamical treatment. Within the
rigid-rotor approximation, these differences involve changes in {\it
  i)} the internal state-averaged geometry, {\it ii)} the center of
mass position, {\it iii)} the reduced mass of the total system, {\it
  iv)} the energy level spacing and {\it v)} the rotation of the
principal axes of inertia. All these effects were considered here for
ND$_2$H and ND$_3$, except the change of internal geometry: the
averaged geometry of NH$_3$ was employed for both deuterated
isotopologues, thereby neglecting the deuterium substitution effect on
the N-H bond length and $\widehat{\rm HNH}$ angle. This assumption was
previously adopted for the NH$_2$D-H$_2$ and ND$_2$H-H$_2$ systems by
\cite{daniel2014} and \cite{wiesenfeld2011}, respectively. The
(rigid-rotor) PES for ND$_3$-H$_2$ was thus again derived from the
CCSD(T) NH$_3$-H$_2$ PES computed by \cite{maret2009}, but in the
principal inertia axes of the isotopologue. We note that internal
geometry effects in the estimate of the rate coefficients are expected to be only moderate 
($\lesssim$ 20-30\%)
at the temperatures investigated here, i.e. $T < 50$K, as demonstrated by
\cite{scribano2010} for the D$_2$O-H$_2$ system.

The transformation between the reference frame of an isotopologue and the
original frame of the NH$_3$-H$_2$ system can be found in
\cite{wiesenfeld2011} (Eqs.~2-3) where the $\gamma$ angle is zero for
ND$_3$ and the coordinates of the ND$_3$ center of mass in the
NH$_3$-H$_2$ reference frame are $X_{\rm CM}$=0.0~ bohr and
$Z_{\rm CM}$=-0.088548436~ bohr (where $X_{\rm CM}$ and $Z_{\rm CM}$
were calculated for the average geometry of NH$_3$ used by
\cite{maret2009}). As for NH$_2$D-H$_2$ and ND$_2$H-H$_2$, the
ND$_3$-H$_2$ PES was generated on a grid of 87~000 points consisting
of 3000 random angular configurations combined with 29 intermolecular
distances in the range 3-15~ bohr. This PES was expanded in products of
spherical harmonics and rotation matrices (see Eq.~4 in Wiesenfeld et
al. 2011), using a linear least-squares fit procedure detailed in
\cite{rist2011}. The fit actually employs the 120--terms expansion
optimized for NH$_3$-H$_2$ by \cite{maret2009}, where details can be
found. The final expansion thus includes anisotropies up to $l_1$=11
for ND$_3$ and $l_2$=4 for H$_2$. The root mean square residual was
found to be lower than 1~cm$^{-1}$ for intermonomer separations $R$
larger than 4.5~ bohr, with a corresponding mean error on the expansion
coefficients smaller than 1 cm$^{-1}$.

\section[]{Collisional dynamics} \label{dynamique}
   
For the various symmetries of ND$_2$H and ND$_3$\footnote{The ND$_2$H
  molecule is an asymmetric top which can be described by the quantum
  numbers $J$, $K_a$ and $K_c$. In what follows, we use the pseudo
  quantum number $\tau$ = $K_a$ -$K_c$ and adopt the notation $J_\tau$
  to describe the rotational structure. The ND$_3$ molecule is a
  symmetric top which is described by the $J$ and  $K$ quantum numbers
  and we hence label the levels as $J_K$. For both molecules, the
  umbrella vibrational motion further splits every rotational level in
  two energy levels. We then introduce the $\epsilon$ quantum number
  in order to distinguish the levels in every doublet. With $\epsilon
  = \pm 1$, the ND$_2$H and ND$_3$ energy levels are then respectively
  labeled as $J_\tau^{\pm}$ and $J_K^{\pm}$ (see details in
  appendix~A).} the calculations aim at providing rate
coefficients up to $T$ = 50~K and for rotational 
energy levels below $\sim$100
cm$^{-1}$. More precisely, in the case of ND$_2$H, we thus provide
rate coefficients for the levels up to $J_\tau = 3_3$. For the para
symmetry of ND$_3$, we consider the levels up to $J_K$ = $4_0$ and
for the ortho symmetry, up to $J_K$ = $4_1$ (see Table \ref{energies_ND3}).

Depending on the deuterated isotopologue, we either solved the
collisional dynamics using the MOLSCAT\footnote{J. M. Hutson and
  S. Green, MOLSCAT computer code, version 14 (1994), distributed by
  Collaborative Computational Project No. 6 of the Engineering and
  Physical Sciences Research Council (UK)} computer code or
HIBRIDON\footnote{HIBRIDON is a package of programs for the
  time-independent quantum treat- ment of inelastic collisions and
  photodissociation written by M. H. Alexander, D. E. Manolopoulos,
  H.-J. Werner, B. Follmeg, Q. Ma and P. J. Dagdigian, with
  contributions by P. F. Vohralik, D. Lemoine, G. Corey, R. Gordon,
  B. Johnson, T. Orlikowski, A. Berning, A. Degli-Esposti, C. Rist,
  B. Pouilly, G. van der Sanden, M. Yang, F. de Weerd, S. Gregurick,
  J. Klos and F. Lique. More information and/or a copy of the code can
  be obtained from the website
  http://www2.chem.umd.edu/groups/alexander/hibridon/hib43.} package
which includes the treatment of both inversion motion and
non--spherical linear collider .  The former was used to treat the
 ND$_2$H-H$_2$  system, since we neglected the inversion motion for
this particular isotopologue. The inversion motion
splits every rotational level in two rotation--inversion energy
levels, the splitting being of the order $\sim$0.4 cm$^{-1}$ for
ND$_2$H. However, the two levels of every doublet are associated to
different ND$_2$H nuclear--spin symmetries, either the ortho or para
species, which are not connected through inelastic collisions. Hence, by
neglecting the inversion motion we consider a system which is suitable
to describe both the ortho and para species, if we omit the small
error (0.4~cm$^{-1}$) made in the estimate of the energy of the
rotation--inversion state.  The methodology of the corresponding
calculations is similar to the recent study of the 
 NH$_2$D-H$_2$ collisional system \citep{daniel2014}. In the case of
ND$_2$H, we followed the description adopted in
\citet{wiesenfeld2011}, which is based on the spectroscopic parameters
given by \citet{coudert1986,delucia1975}.  Hence, we described the
energy levels adopting the rotational constants A = 5.3412, B = 7.4447
and C= 3.7533 cm$^{-1}$ and the centrifugal terms $D_{JJ}$ = $3.346 \,
10^{-4}$, $D_{JK}$ = $-4.931 \, 10^{-4}$ and $D_{KK} $ = $2.164 \,
10^{-4}$, these values being given in the II$^{r}$ representation \citep{gordy1984}
suitable to MOLSCAT.

 In the case of ND$_3$, two levels of a doublet can
pertain to the same symmetry and thus can be connected through
collisions. The MOLSCAT code could be used to treat such a problem, but
only in the case of a spherical collider, like He or p--H$_2$
restricted to its fundamental rotational level $J_2=0$. Since our
calculations aim at including the first excited state of p--H$_2$
(i.e. $J_2$=2), we used the HIBRIDON package which includes the
possibility of treating the inversion motion in the case of a
non--spherical collider. We checked the consistency of the results
obtained with the two codes, for the collisions of o--ND$_3$ with
H$_2$ ($J_2=0$) and we could verify that the cross sections agree with
an accuracy typically of the order of the per cent.
\begin{table}
\caption{Theoretical energy levels of the three ND$_3$ symmetries, for the first 20 energy levels. For each symmetry, we give the quantum numbers
$J_K^\epsilon$ and the energy in cm$^{-1}$. }
\begin{center}
\begin{tabular}{cccccc}
\multicolumn{2}{c}{o--ND$_3$} &  \multicolumn{2}{c}{p--ND$_3$}  & \multicolumn{2}{c}{m--ND$_3$} \\ \hline
   $1_1^-$  &    8.27 & $0_0^+$  &    0.05 & $0_0^+$  &    0.00 \\ 
   $1_1^+$  &    8.32 & $1_0^+$  &   10.29 & $1_0^+$  &   10.34 \\ 
   $2_2^+$  &   22.78 & $2_0^+$  &   30.91 & $2_0^+$  &   30.86 \\ 
   $2_2^-$  &   22.83 & $3_3^+$  &   43.55 & $3_3^-$  &   43.55 \\ 
   $2_1^+$  &   28.84 & $3_3^-$  &   43.60 & $3_3^+$  &   43.60 \\ 
   $2_1^-$  &   28.89 & $3_0^+$  &   61.71 & $3_0^+$  &   61.76 \\ 
   $3_2^-$  &   53.64 & $4_3^-$  &   84.69 & $4_3^+$  &   84.69 \\ 
   $3_2^+$  &   53.69 & $4_3^+$  &   84.74 & $4_3^-$  &   84.74 \\ 
   $3_1^-$  &   59.69 & $4_0^+$  &  102.90 & $4_0^+$  &  102.85 \\ 
   $3_1^+$  &   59.74 & $5_3^+$  &  136.12 & $5_3^-$  &  136.12 \\ 
   $4_4^+$  &   70.56 & $5_3^-$  &  136.17 & $5_3^+$  &  136.17 \\ 
   $4_4^-$  &   70.61 & $6_6^-$  &  143.34 & $6_6^+$  &  143.34 \\ 
   $4_2^+$  &   94.78 & $6_6^+$  &  143.39 & $6_6^-$  &  143.39 \\ 
   $4_2^-$  &   94.83 & $5_0^+$  &  154.28 & $5_0^+$  &  154.33 \\ 
   $4_1^+$  &  100.84 & $6_3^-$  &  197.83 & $6_3^+$  &  197.83 \\ 
   $4_1^-$  &  100.89 & $6_3^+$  &  197.88 & $6_3^-$  &  197.88 \\ 
   $5_5^-$  &  103.83 & $7_6^+$  &  215.34 & $7_6^-$  &  215.34 \\ 
   $5_5^+$  &  103.88 & $7_6^-$  &  215.39 & $7_6^+$  &  215.39 \\ 
   $5_4^-$  &  121.99 & $6_0^+$  &  216.04 & $6_0^+$  &  215.99 \\ 
   $5_4^+$  &  122.04 & $7_3^+$  &  269.83 & $7_3^-$  &  269.83 \\  \hline
\end{tabular}
\end{center}
\label{energies_ND3}
\end{table}%
To define the ND$_3$ energy structure, we adopted the rotational constants
$A=B=5.142676$ and $C = 3.124616$ cm$^{-1}$ and set the inversion splitting at a constant value 
of 0.05 cm$^{-1}$. The corresponding energy levels are given in Table \ref{energies_ND3}.
As can be seen in this table, the energy levels
associated with the para and meta symmetries of ND$_3$ are mostly
identical. The only differences between the two sets of energies are
found for the  $K$=0 levels. This similarity entails that the cross
sections associated to the two symmetries will be roughly identical
for a given $J_K^{\epsilon} \to {J'}_{K'}^{\epsilon '}$ transition.
Indeed, the coupling terms that enter the dynamic equations are identical
for the two symmetries and the only differences in the two sets of equations will thus
come from the differences in the energy levels.
These similarities are verified in our state--to--state cross sections
performed at a few energies.  Cross sections involving levels with
both initial and final $K \neq 0$ differ at most by 1\%. For
transitions that involve levels with $K=0$, the differences may be
higher but never exceed 5\%. Hence, we chose to perform
calculations for only one symmetry, i.e. for p--ND$_3$. Full details
on the spectroscopy of ND$_3$ are presented in Appendix~A.

The accuracy of the cross sections is assessed by a few test
calculations where we checked the convergence against the step of
integration used to propagate the wavefunctions, and the number of
energy levels considered during the propagation. These parameters are
summarized in Table \ref{table:conv_ND2H} for ND$_2$H and Table
\ref{table:conv_ND3} for ND$_3$.  In the case of ND$_3$, the
integration step is not reported since we used a single value of 0.1
$a_0$ at every energies. The parameters quoted in these tables ensure a
convergence better than 5\% at all energies. In these tables, we
also report the step between two consecutive energies. 
These values ensure a good characterization of the resonances, for the
transitions between levels under consideration.  For both
ND$_2$H and ND$_3$, all the calculations were performed within the
close--coupling formalism.
Recently, \citet{tkac2015} reported quantum dynamical calculations for
the ND$_3$-H$_2$ collisional system, with the aim of testing the PES
against experimental differential cross sections. These comparisons
were performed at total energies above 430 cm$^{-1}$. They assessed
the convergence of their calculations with respect to the size of the
rotational basis of the two rotators and, in particular, they found
that omitting the $J_2 = 4$ energy level of p--H$_2$ can lead to
misestimate the cross sections by $\sim$25\%, for transitions
associated with the $J_2 = 0 \to 0$ H$_2$ state--to--state
transitions.  Hence, we checked the accuracy of our calculations by
testing the influence of the size of the H$_2$ rotational basis, at
various total energies between 200 and 400 cm$^{-1}$. We find that
below 400 cm$^{-1}$, the cross sections were affected by less than 5\%
due to the truncation of the p--H$_2$ rotational basis to $J_2=0, 2$. At
400 cm$^{-1}$, we indeed found that some transitions were more largely
affected, with inaccuracies of up--to $\sim$35\%. However, less than
5\% of the total number of transitions are affected by errors greater
than 5\%. Moreover, the corresponding cross sections are the lowest in
magnitude, i.e. lower than 0.001 $\AA^2$, the highest cross sections
being of the order of 10 $\AA^2$ at this total energy. Hence, given
the low influence of the $J_2 = 4$ energy level of p--H$_2$ for the
range of energy considered in the current work, and since the
inclusion of this level would considerably raise the calculation time,
we chose to restrict the rotational levels of H$_2$ to $J_2$ = 0, 2 .
   
\begin{table}
\caption{Parameters that describe the MOLSCAT calculations performed
  for  ND$_2$H-H$_2$, i.e. the step between the consecutive total
  energies used to characterize the cross sections (second column),
  the STEPS parameter (third column) and the size of the rotational
  basis (fourth column).}
\begin{center}
\begin{tabular}{cccc}
\hline
 Energy range (cm$^{-1}$) & step in energy (cm$^{-1}$) & STEPS & JMAX \\ \hline
         $<$ 25  & 0.1 & 40 &  7 \\
    25 $-$  60  & 0.1 & 20 &  7 \\
    60 $-$  80  & 0.1 & 10 &  7 \\
    80 $-$ 110 & 0.1 & 10 &  8 \\
  110 $-$ 155 & 0.2 & 10 &  8 \\
  155 $-$ 250 & 0.5 & 10 &  8 \\
  250 $-$ 400 & 1    & 10 &  8 \\ \hline
\end{tabular}
\end{center}
\label{table:conv_ND2H}
\end{table}%

\begin{table}
\caption{Parameters that describe the HIBRIDON calculations performed for the ortho and para symmetries of  ND$_3$-H$_2$, i.e. the step between the consecutive total energies used to characterize the cross sections (second column), and the size of the rotational basis given by the highest rotational quantum number (third column) and the cut in energy (fourth column).}
\begin{center}
\begin{tabular}{cccc}
\hline
 Energy range (cm$^{-1}$) & step in energy (cm$^{-1}$) & JMAX & E$_{cut}$\\ \hline
         $<$ 30  & 0.1 &   6   &  +$\infty$ \\
    25 $-$  65  & 0.1 &   7   &  +$\infty$ \\
    65 $-$  105    & 0.1 &   8   &   650 \\
  105 $-$  150    & 0.1 &   9   &   700 \\
  150 $-$  200    & 0.1 &   9   &   750 \\
  200 $-$  300    & 0.5 &  10   &   800 \\
  300 $-$  450    & 2.0 &  10   &   800 \\ \hline
\end{tabular}
\end{center}
\label{table:conv_ND3}
\end{table}%

\section{Rate coefficients} \label{rates}

The de-excitation rate coefficients for the rotation-inversion
transitions were calculated by averaging the above cross sections with
a Maxwell--Boltzmann distribution that describes the distribution of
velocity of the molecules in the gas \citep[see e.g. eq. (5)
  in][]{wiesenfeld2011}. The rate coefficients were thus obtained for
temperatures in the range 5--50~K and for all transitions involving
NHD$_2$ levels below 82~cm$^{-1}$ (i.e. up to $J_\tau=3_3$) and ND$_3$
levels below 103~cm$^{-1}$ (i.e. up to $J_K=4_0$ and $4_1$ for 
the para and ortho species, respectively).

Since the hyperfine structure due to the nitrogen nuclei ($^{14}$N) is
resolved observationally (see Section~6), rate coefficients for the
hyperfine transitions are also required. The recoupling theory now
routinely employed for linear species
\citep[see e.g.][]{daniel2004,daniel2005,kalugina2015} has not yet been extended to
non-linear molecules for which only the statistical
approximation is possible. In this approach, it is assumed that the
hyperfine de-excitation rate coefficients are proportional to the
degeneracy (2$F'$ + 1) of the final hyperfine level and completely
independent of the initial hyperfine level. This corresponds to a
statistical reorientation of the quantum number $F$ after collision
and it is also called the $M_j$ randomizing limit or proportional
approach. It has been shown to be inaccurate by large factors when
predicting hyperfine individual rate coefficients
\citep[see e.g.][]{daniel2005,faure2012}. On the other hand, when the opacity of
the individual hyperfine lines is low ($\sim$1), the hyperfine levels are populated
according to their statistical weights and, in this regime, the
statistical approximation is well adapted \citep{daniel2005,faure2012}. As
the hyperfine transitions of the deuterated isotopologues are usually
optically thin, the statistical approximation should be accurate
enough to model the spectra of NH$_2$D, NHD$_2$ and ND$_3$.
Note that in radiative transfer models, the quasi--elastic $J, F \to J, F'$ rate 
coefficients are also needed. Unfortunately, these rate coefficients cannot be guessed from elastic 
rate coefficients, which are typically of the order $10^{-9}$ cm$^{3}$ s$^{-1}$. 
Indeed, \citet{faure2012} showed that applying the proportional approach 
to elastic rate coefficients lead to overestimate the quasi--elastic rates by large factors, 
with respect to rate coefficients calculated with a recoupling method. Hence, in the radiative transfer
calculations described in Sect. 6, we applied the proportional approach to a canonical rate of 
$10^{-10}$ cm$^{3}$ s$^{-1}$ in order to obtain an estimate of the quasi--elastic rate coefficients.

The whole set of hyperfine rate coefficients will be available through
the LAMDA \citep{schoier2005} and BASECOL \citep{dubernet2013}
databases.

\section{Discussion} \label{discussion}

To date, two sets of rate coefficients were computed for ND$_2$H.  The
first set considered He as a collider \citep{machin2007}, while the
second set was calculated for p--H$_2$ using a $J_2=0$ basis set
\citep{wiesenfeld2011}. Quite often, the collisions with He are used
to emulate the rate coefficients with p--H$_2$. However, as for other
molecular systems, the comparison between these two sets showed that
the differences can be quite large, the rates with p--H$_2$($J_2$=0)
being higher by a factor 3--30 depending on the transition.

\begin{figure}
\begin{center}
\includegraphics[angle=0,scale=0.4]{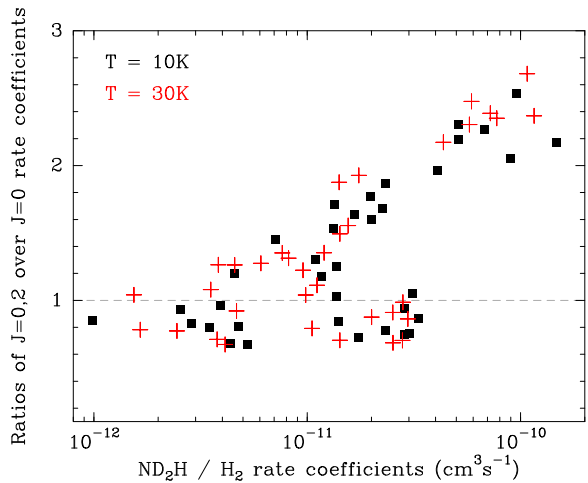}
\caption{Ratio between the ND$_2$H / H$_2$ rate coefficients calculated in the current work
with a J=0,2 basis for H$_2$, with those of \citet{wiesenfeld2011} that were obtained with a J=0 basis. The comparison
considers the levels up--to $J_\tau = 2_2$ and is performed at T = 10K (black points) and T = 30K (red crosses).}
\label{comp_Wiesenfeld}
\end{center}
\end{figure}

In order to perform some comparison with these previous works, we
considered the  ND$_2$H-H$_2$ rate coefficients determined by
\citet{wiesenfeld2011}, where data were computed for the nine first
rotational energy levels and for temperatures in the range 5--35K.
These calculations only differ from the current ones in two
points. First, in \citet{wiesenfeld2011}, the cross sections
were computed in part with the coupled--states approximation.  In the current
study, all the calculations were performed within the close--coupling
formalism. Secondly, the present H$_2$ rotational basis includes the
$J_2=2$ level, while it was reduced to $J_2=0$ in
\citet{wiesenfeld2011}. As already commented in \citet{daniel2014},
increasing the H$_2$ basis can have a non negligible effect on the
estimate of the magnitude of the rate coefficents. In the case of
NH$_2$D, it was found that using a $J_2=0$ basis rather than a
$J_2=0,2$ basis can lead to underestimating by a factor $\sim$3 the
cross sections of highest magnitude \citep[see Fig. 2
  in][]{daniel2014}. Moreover, in that study, it was already mentioned
that the rates given by \citet{wiesenfeld2011} should be accurate to
within a factor $\sim$3. Such a factor was guessed from a few
preliminary calculations and considering the behaviour of a reduced
set of cross sections \citep[see Fig. 9 in][]{daniel2014}.  With the
current data, the comparison can be extended and done on more
quantitative grounds. In Fig. \ref{comp_Wiesenfeld}, we show the ratio
between the two sets of rate coefficients as a function of the
magnitude of the rate coefficients. From this figure, it can be seen
that the largest differences are found for the rates of highest
magnitude. This is similar to what was already observed in the 
NH$_2$D-H$_2$ collisional system. Additionally, the alteration of the rates
of highest magnitude is at most of a factor $\sim$2.5, which is of
the same order of magnitude as for the NH$_2$D-H$_2$ collisional system.

\begin{figure}
\begin{center}
\includegraphics[angle=0,scale=0.4]{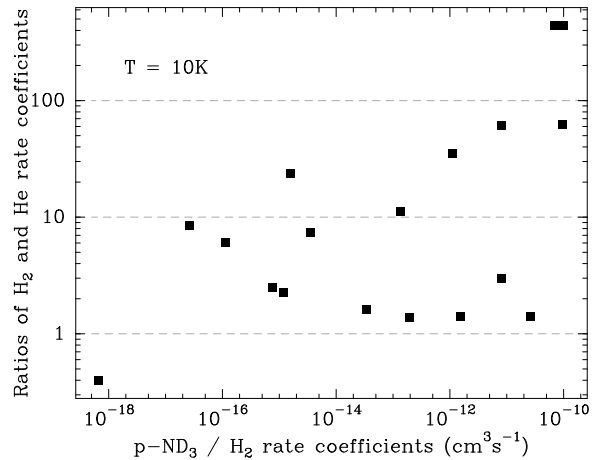}
\caption{
  Ratio of the H$_2$ over He rate coefficients of p--ND$_3$. The
  comparison considers the levels up--to $J_K^{\epsilon} = 4_3^-$ and
  is performed at T = 10K.}
\label{comp_Machin}
\end{center}
\end{figure}

The calculation of  ND$_3$-He rate coefficients was a goal
of the PhD thesis of L. Machin
\citep{machin2006}\footnote{https://tel.archives-ouvertes.fr/tel-00128133}
(these unpublished rate coefficients are given in Table 5.3 of the
thesis manuscript).  As already evidenced for the NH$_2$D and NHD$_2$
isotopologues, the differences between the He and H$_2$ rate
coefficients are large \citep[see Fig. 8 in][]{daniel2014}. In
Fig. \ref{comp_Machin}, we plot the ratio of the H$_2$ and He
rate coefficients, at a temperature $T$=10K and for the levels up to
$J_k^{\epsilon} = 4_3^-$. As for NH$_2$D and NHD$_2$, the ratios are in
the range 1--100 and the largest differences are obtained for the
highest rate coefficients. Interestingly, the calculations of Machin
with He were performed for both the para and meta symmetries
(however, the meta symmetry was misleadingly referenced as ortho in
that work).  Machin could thus check that the rate coefficients for the
para and meta species agree with a good accuracy, as expected. In
fact, he found that at temperatures higher than 50K, the agreement is
better than a few per cent.  The comparison at $T$=10K shows larger
differences, but most of the rate coefficients agree  to within 20\% (in
Table 5.3 of the manuscript, the $1_0^+ \to 3_3^-$ transition is the
only transition that shows a higher ratio, i.e. $\sim$ 2). These
differences in the para and meta rate coefficients were attributed
to differences in the resonance structure.  In
Sect. \ref{dynamique}, we mentioned that we performed some test
calculations at some specific energies and we could verify that at
these energies, the meta and para cross sections agree within a
few per cent. However, as in the case of the collisions with He, we
cannot discard that the resonance structure for the two symmetries
could induce larger differences, but not above $\sim$20\%, which is
the typical accuracy of the present rate coefficients.

\begin{figure}
\begin{center}
\includegraphics[angle=0,scale=0.3]{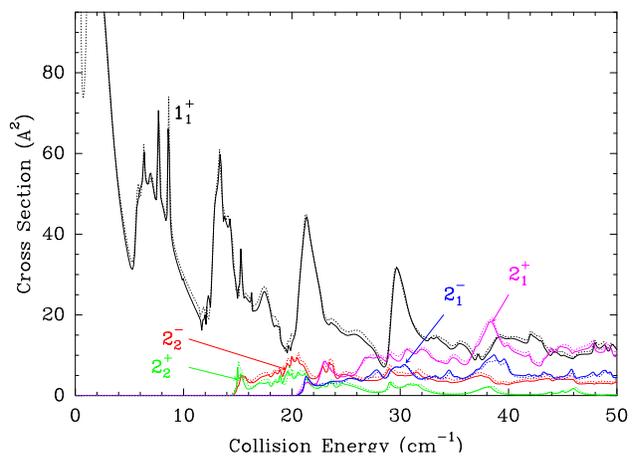}
\caption{Cross sections for o--ND3 that originate from the $1_1^-$ level. This figure can directly be compared
to Fig. 13 of \citet{ma2015} and the cross sections obtained in the latter study are reported as dotted lines. }
\label{comp_Ma}
\end{center}
\end{figure}

Recently, \citet{ma2015} reported dynamical calculations for the
ND$_3$-H$_2$ system that were performed on the same PES as the
current calculations. These two sets of calculations were performed
independently and with a different focus. While in the current case,
we are interested in providing collisional rate coefficients,
\citet{ma2015} put emphasis on the description of the resonance
features in order to probe the possibility of characterizing these
resonances in future experiments. Since the energy, strength and shape
of the resonances depend strongly on the accuracy of the PES, this
would provide a test of the accuracy of the PES of \cite{maret2009}
and hence, would give an estimate of the accuracy we could expect for
the rate coefficients. In Fig. \ref{comp_Ma}, we give the cross
sections of o--ND$_3$ that originate from the $1_1^-$ level.   
This figure is similar to Fig. 13 of \citet{ma2015} and in the current figure, the cross
sections obtained by \citet{ma2015} are reported as dotted lines.  The overall
agreement between the two sets of calculations is very good, even if
there remains some small differences in the description of the
resonances. These differences do not exceed a few per cent and
presumably originate from the respective accuracy of the two collision
dynamics calculations .

\section{Astrophysical modeling} \label{astro}

The rate coefficients described in the previous sections are a key
ingredient for astrophysical models interpreting the deuterated
ammonia inversion/rotational lines observed from far--infrared to
centimeter wavelengths. Such observations of the NH$_2$D, ND$_2$H and
ND$_3$ isotopologues were reported and analyzed by
\citet{tine2000,roueff2000,saito2000,lis2002,vandertak2002,roueff2005,lis2006,gerin2006},
but the lack of rate coefficients hampered the analysis, which thus
relied on simplifying assumptions.  In some cases, the recourse to
such simplification leads to discrepancies, as outlined by
\citet{lis2006}. Indeed, it was then shown that an LTE analysis
applied to two rotational lines of o--ND$_2$H, at 336 and 389 GHz,
gives inconsistent column density estimates. In the case
of the observations towards the prestellar cores 16293E or B1b, it was
found that the two estimates disagree by factors of $\sim$3 and 6,
respectively.

In what follows, we reconsider the observations available towards B1b
and 16293E, in light of the newly calculated ND$_2$H and ND$_3$
rate coefficients presented here, as well as the NH$_2$D / H$_2$ rate
coefficients described in \citet{daniel2014}. The molecular excitation
and radiative transfer is solved with the \texttt{1Dart} code
described in \citet{daniel2008}. As in \citet{daniel2013}, we report column
densities, which are calculated by counting the number of molecules across the diameter of the spheres
that describe the sources. These column densities are not tied to any particular molecular transition
and hence or not averaged over a particular telescope beam size. Such beam averaged column densities are
 often introduced when estimating
molecular column densities and in what follows, we will sometimes introduce it  for the sake of comparison with previous results.

\subsection{Barnard 1b}

B1b, located at 235 pc in the Perseus molecular cloud, is a region of active low--mass
star formation. The region around B1b has been the subject of many observational studies \citep[see references in][]{daniel2013}
and consists of two objects, B1b--N and B1b--S \citep{hirano1999}, whose evolutionary states are still under debate.
Currently, B1b--N is thought to be in the first hydrostatic core stage while B1b--S seems 
to be a slightly more evolved object \citep{hirano2014,gerin2015}.

\subsubsection{NH$_2$D} 

\begin{figure*}
\begin{center}
\includegraphics[angle=0,scale=0.6]{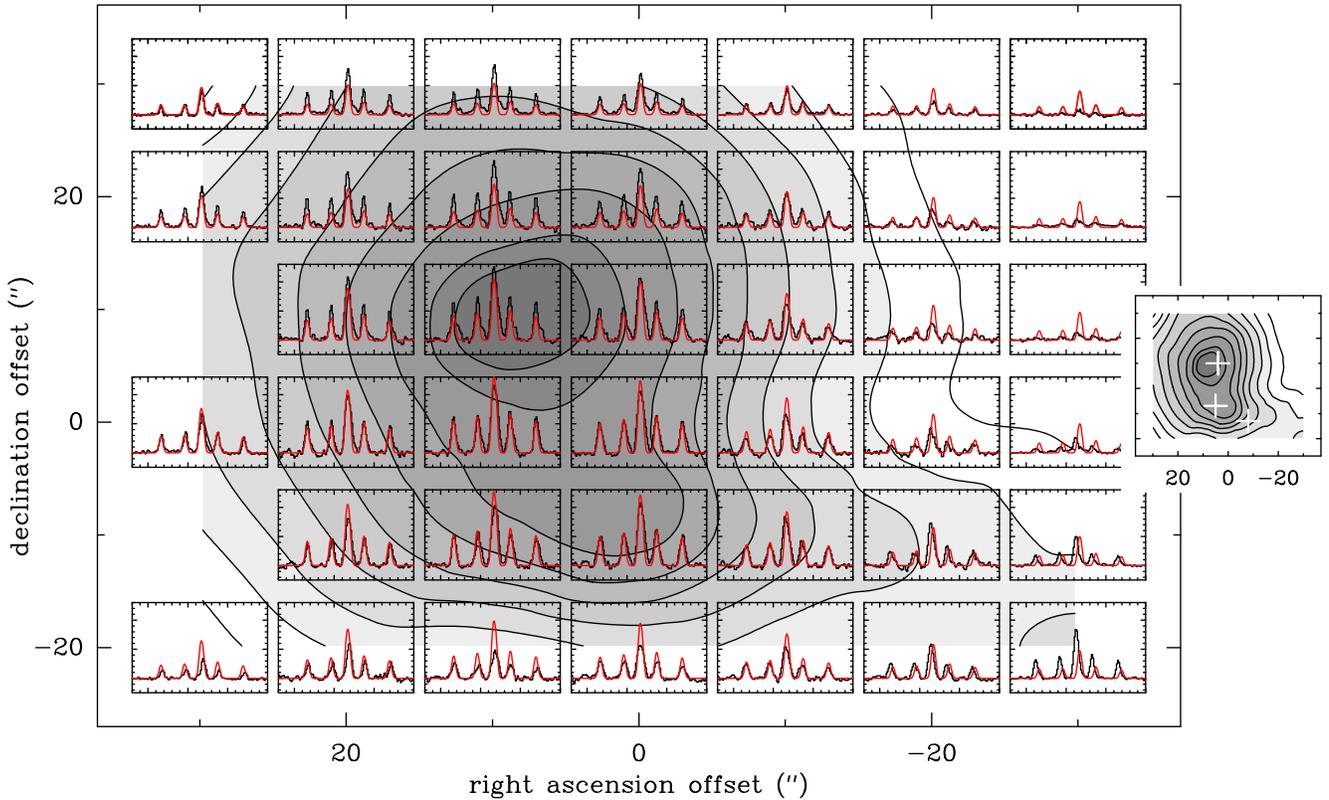}
\caption{Comparison between model (red curves) and observations (histograms) for the o--NH$_2$D $1_{1,1}$--$1_{0,1}$ line towards B1b.The inset map on the right shows the isocontours with the same scale for the right ascension and declination. The white crosses indicate the position of the B1-bN and B1-bS sources identified by \citet{hirano1999}, as well as the Spitzer source reported by \citet{jorgensen2006}.}
\label{NH2D_map_B1} \vspace{-0.1cm}
\end{center}
\end{figure*}

\begin{figure*}
\begin{center}
  \subfigure[]{\includegraphics[angle=0,scale=0.3]{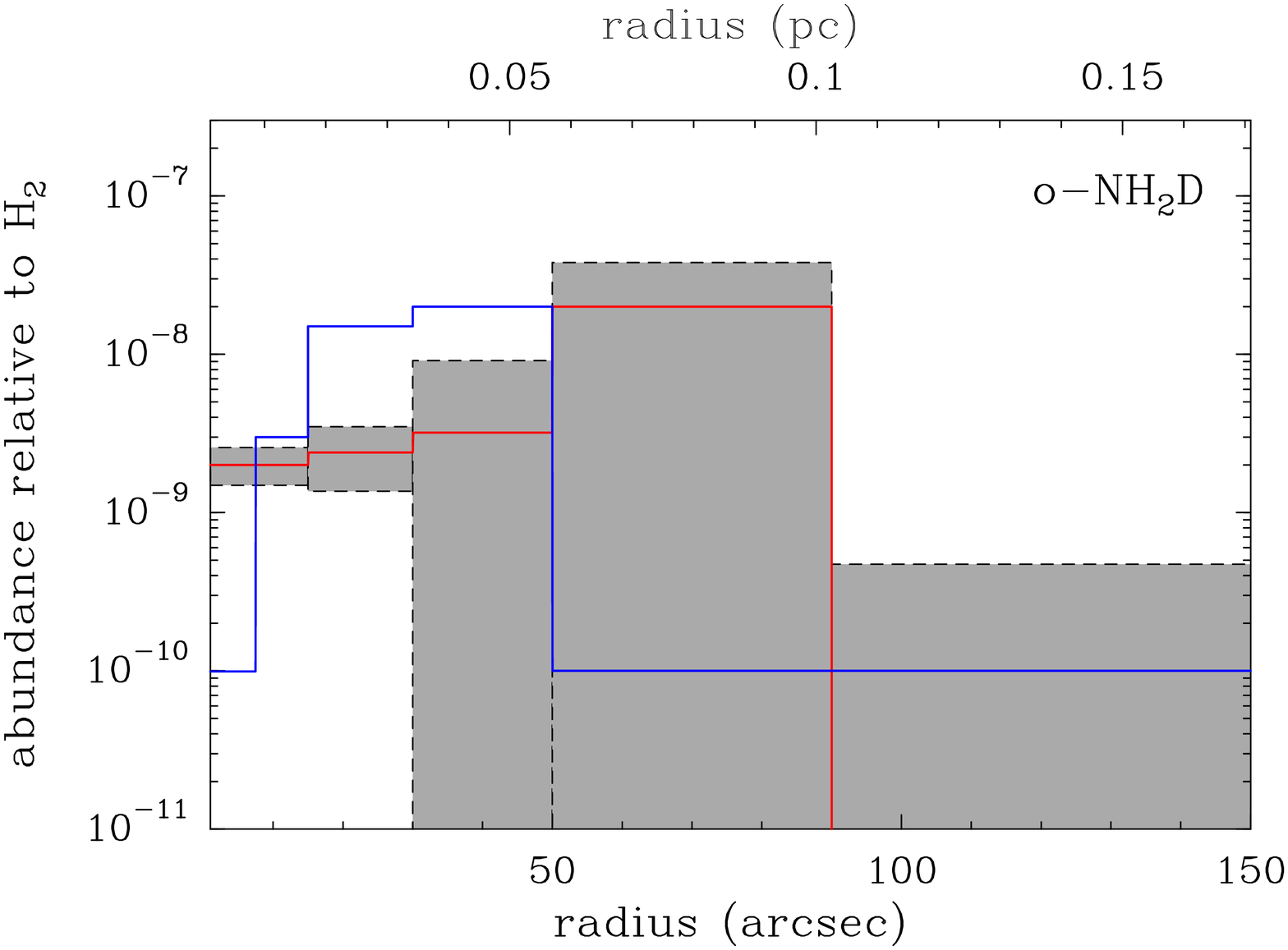}} \quad
  \subfigure[]{\includegraphics[angle=0,scale=0.3]{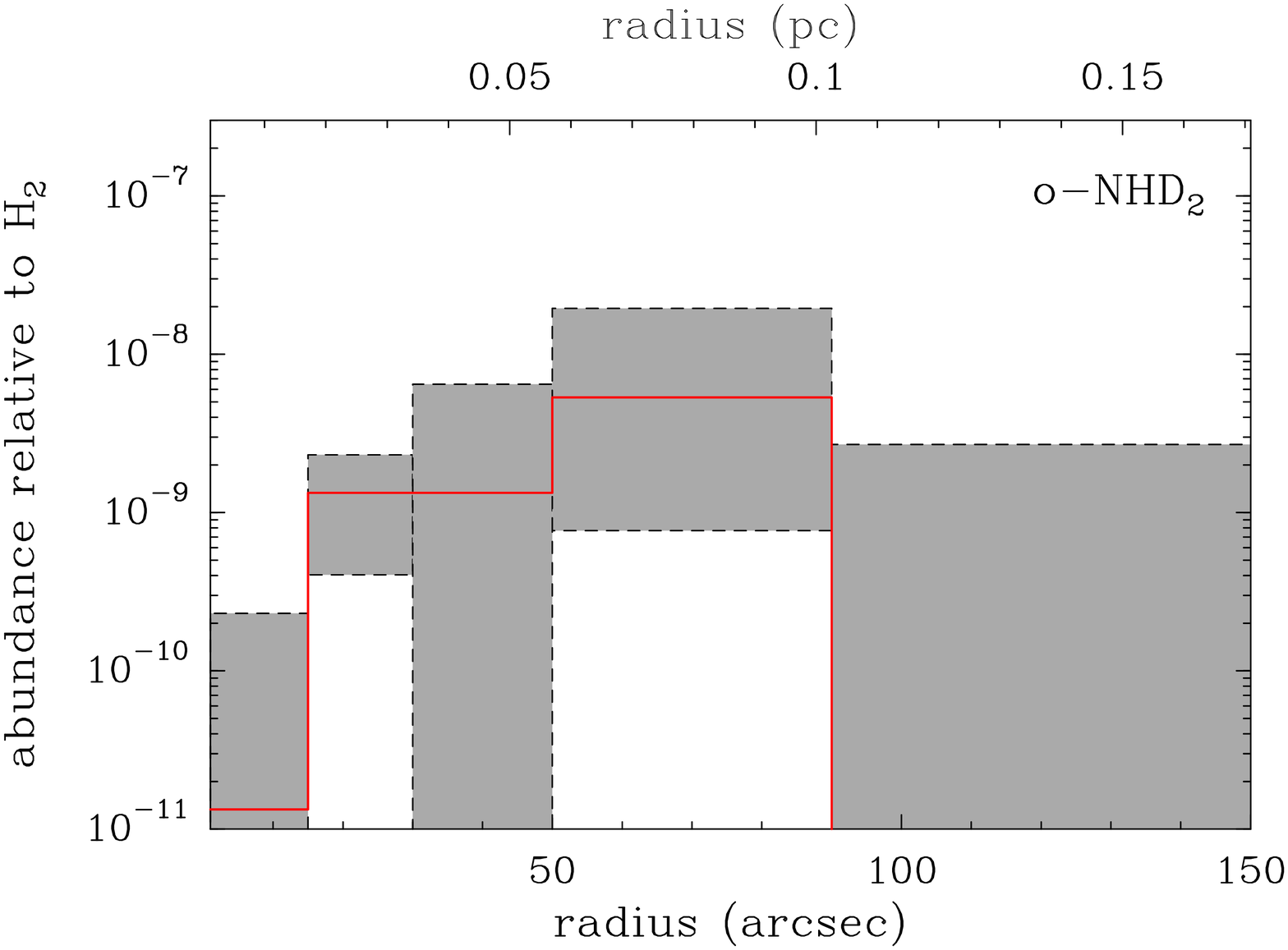}} \\
\end{center}
\caption{ \textbf{(a)} 
  Abundance profile of o--NH$_2$D (red curve) as a function of
  radius in the prestellar core B1b. The grey areas show the
  confidence zone for the abundance at every radius. The abundance
  profile derived by \citet{daniel2013} is indicated as a blue curve.
  \textbf{(b)}  Abundance profile of o--NHD$_2$ (red curve) as a function of
  radius.}
\label{profils_B1}
\end{figure*}

\begin{figure*}
\begin{center}
\subfigure[]{\includegraphics[angle=0,scale=0.3]{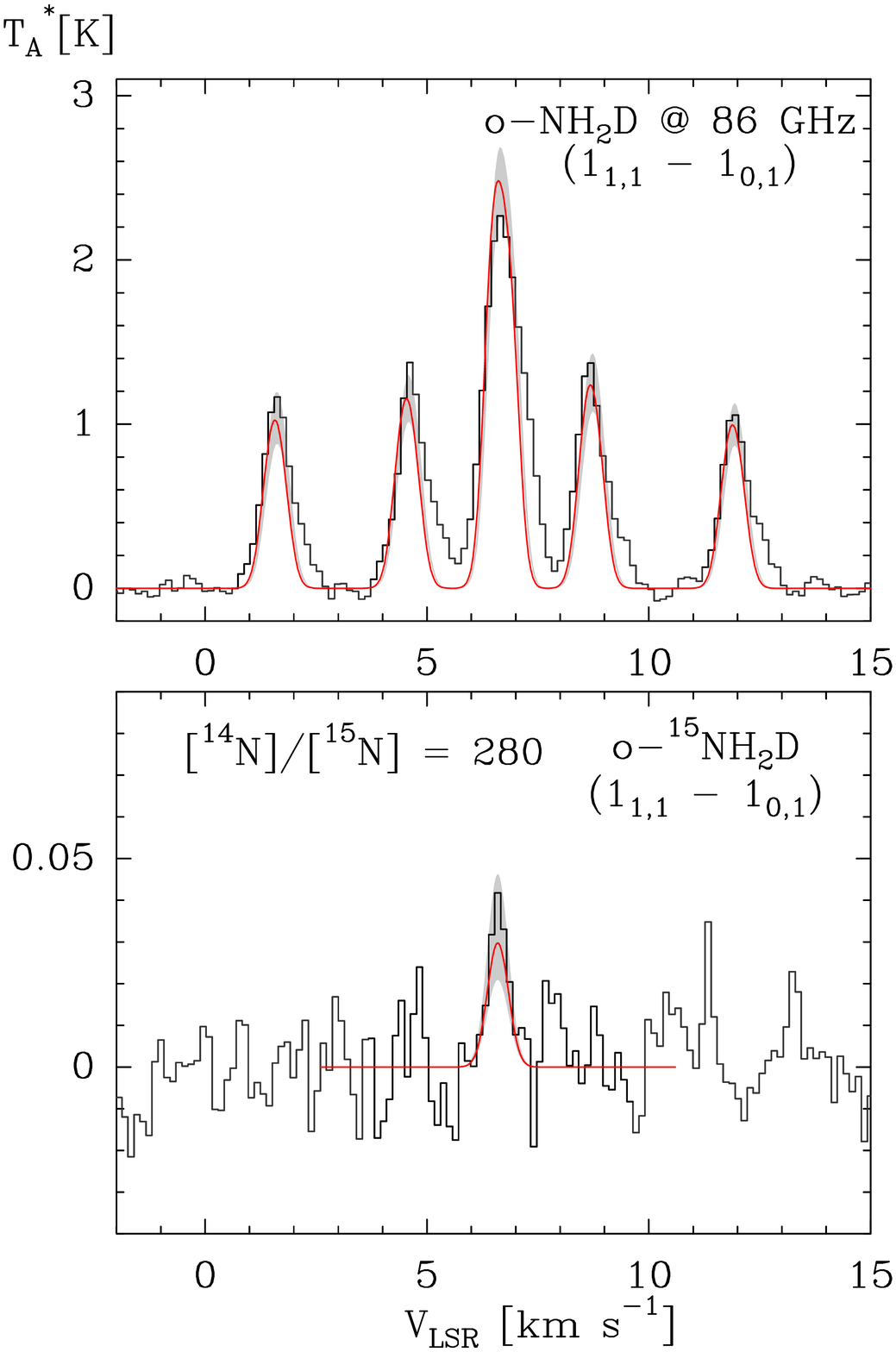}} \quad
\subfigure[]{\raisebox{10mm}{\includegraphics[scale=0.4]{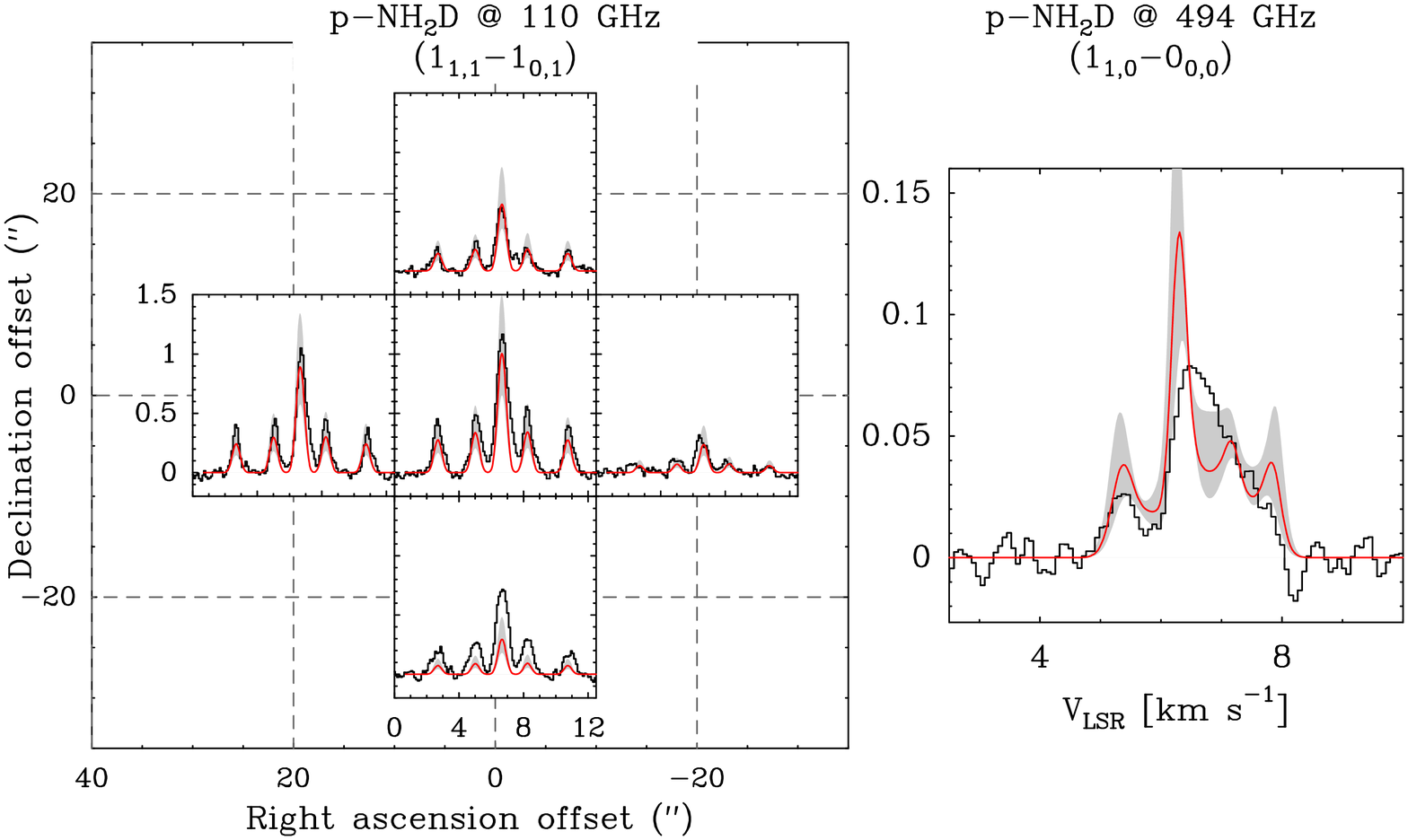}}}
\end{center}
\caption{
Comparison between model (red curves) and observations (histograms), in the T$_A^*$ scale, for the NH$_2$D lines
observed towards B1b:
\textbf{(a)} o--NH$_2$D and o--$^{15}$NH$_2$D $1_{1,1}$--$1_{0,1}$ lines
\textbf{(b)} mini--map of the p--NH$_2$D $1_{1,1}$--$1_{0,1}$ line (left panel) and single point observation of the $1_{1,0}$--$0_{0,0}$ line (right panel) 
}
\label{fig:NH2D}
\end{figure*}

The physical structure adopted for this source (i.e. dust and gas
temperatures, H$_2$ density) is taken from \citet{daniel2013}, where
it was inferred from the analysis of 850 $\mu$m and 1.3 mm continuum observations.
In the latter study, the map of the o--NH$_2$D
$1_{1,1}$--$1_{0,1}$ emission was already analyzed by solving the
molecular excitation. However, the rate coefficients for the
NH$_2$D--H$_2$ system were not available at that time and the
analysis thus relied on the NH$_2$D--He rate coefficients \citep{machin2006},
that were subsequently scaled. We refer the reader to the discussion
in \citet{daniel2013} for the details that concern the scaling of the
rates. More recently, \citet{daniel2014} reported rate coefficients
for the NH$_2$D--H$_2$ system. Using these new rate coefficients with
the o--NH$_2$D abundance profile given in \citet{daniel2013}, we find
that it leads to overestimate the line
intensity by a factor $\sim$2. This illustrates the inaccuracy introduced by the scaling of
the rate coefficients.
We thus re--analyze the o--NH$_2$D and o--$^{15}$NH$_2$D data
available towards B1b with the same methodology as in \citet{daniel2013}. 
In particular, we note that the B1b is a complex 
region, with two cores separated by $\sim$20'', which have V$_{LSR}$ that differ 
by $\sim$1 km s$^{-1}$. Hence, to model the spectra, we follow the methodology described 
in \citet{daniel2013}: we use a 1D spherical model, adopt a 
V$_{LSR}$$\sim$6.7 km s$^{-1}$ and constrain the models by fitting the blue part of the spectra.
The estimate of the NH$_2$D abundance profile
across the core is constrained thanks to the o--NH$_2$D
$1_{1,1}$--$1_{0,1}$ emission map and the comparison between the model
and observations is given in Fig. \ref{NH2D_map_B1}. 
The corresponding o--NH$_2$D abundance profile is reported in red in
Fig. \ref{profils_B1}. The confidence zone, as defined in \citet{daniel2013},
is indicated as grey area.
In this figure, the abundance profile derived by \citet{daniel2013} is shown in blue.
Finally, the o--$^{15}$NH$_2$D data observed towards the central
position was also re-analyzed. The comparison of the models and
observations for the two isotopologues at this position are given in
Fig.~\ref{fig:NH2D}.

The previous estimate for the o--NH$_2$D column density made by
\citet{daniel2013} was log($N$) = $14.73^{+0.12}_{-0.08}$. With the
new rate coefficients, this value is revised to log($N$) =
$14.76^{+0.10}_{-0.14}$.  For o--$^{15}$NH$_2$D,
the column density was previously estimated to log($N$) =
$12.37^{+0.33}_{-0.11}$ while the current estimate is
$12.31^{+0.20}_{-0.16}$.  The corresponding isotopologue column
density ratio, previously estimated to $230^{+105}_{-55}$ is now 
$280^{+120}_{-100}$.

In the case of the p--NH$_2$D spin isomer, the abundance was constrained
using a mini--map of the $1_{1,1}$--$1_{0,1}$ line at 110 GHz, observed at the IRAM 30m telescope, and the 
 $1_{1,0}$--$0_{0,0}$ line at 494 GHz, observed with Herschel. 
These latter data were obtained using the HIFI instrument \citep{degraauw2010} on Herschel \citep{pilbratt2010}, in the context of the open time program OT2\_dlis\_3 (Herschel observation ID 1342248917). The data were taken in the frequency-switching mode, were processed using the standard HIFI data reduction pipeline and the resulting spectra were subsequently reduced using the GILDAS CLASS software package. The FWHM HIFI beam size is $\sim$44$\arcsec$ at 492 GHz and the main beam efficiency is 62\%\footnote{http://herschel.esac.esa.int/twiki/pub/Public/HifiCalibrationWeb/\-HifiBeamReleaseNote\_Sep2014.pdf}.

\begin{figure}
\begin{center}
\includegraphics[angle=0,scale=0.35]{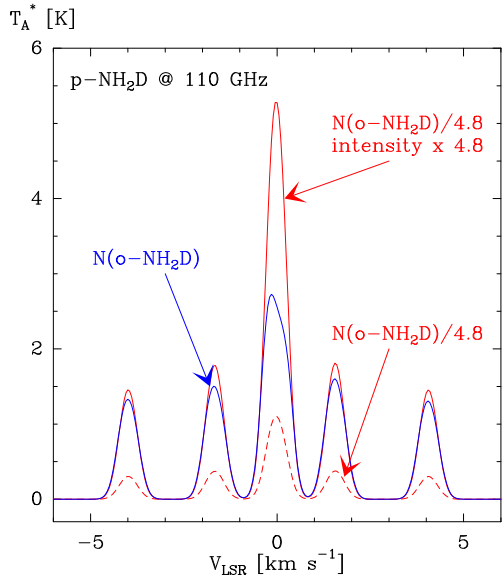}
\caption{Models for the p--NH$_2$D 110 GHz transition. The blue curve gives the intensity we would 
obtain with a p--NH$_2$D abundance profile identical to that of o--NH$_2$D. The red dashed curve corresponds to the model
adopted to reproduce the p--NH$_2$D 110 and 494 GHz lines, with an ortho--to-para ratio of 4.8. The red curve corresponds to the same model with the intensity multiplied by a factor 4.8.}
\label{OTP_NH2D} 
\end{center}
\end{figure}

 By using an overall scaling of the o-NH$_2$D
 abundance profile, we managed to obtain a reasonable fit of the p-NH$_2$D observations, although the line shape of the $1_{1,0}$--$0_{0,0}$ transition is not correctly reproduced. Indeed, the best fit model predicts self--absorption features which are not observed
 (see right panel in Fig. \ref{fig:NH2D}). Some alternative models
 were tried in order to improve the fit, but we could not find a model that would considerably improve the comparison
 with the observations. We thus kept the simplest model, i.e. the model where the p--NH$_2$D abundance is 
 scaled from the o--NH$_2$D abundance profile. The OPR derived from the modeling is $4.8^{+3.2}_{-2.0}$.
 This ratio is higher than the ratio predicted through statistical considerations, i.e. 3, but, compatible given the large 
 error bars. Additionally, we note that examining the possible scenarios relative to the NH$_2$D formation, a ratio of 3
 would be expected in the case of formation on grains (with T$_g$ $\geq$ 10K) while a ratio $<$3 is expected if the formation occurs
 in the gas phase \citep{sipila2015}. Note, however, that it was shown by \citet{persson2015} that, in the case
 of NH$_2$,  the interconversion between ortho and para states due to reactive collisions with H atoms is an efficient process
 that may substantially alter the NH$_2$ OPR. Such process is, however, generally not included in the gas-phase formation models
 due to the lack of either theoretical or experimental value for the reactive rate coefficients.
Finally, to date, OPR higher than 3 are not expected for NH$_2$D. Although \citet{shah2001}
reported similar values, we believe that the  accuracy of the analysis is
influenced by the high opacity of the line, which introduces uncertainties in the estimate of the abundance profile.
This is illustrated by the models shown in Fig. 
 \ref{OTP_NH2D}. In this figure, we show a model for the 110 GHz line where the p--NH$_2$D abundance profile
 is set identical to the o--NH$_2$D profile (blue curve). The corresponding intensities are compared to the  
 model adopted for B1b (red dashed curve). The fact that the intensity does not scale linearly with the column density
 is emphasized through the comparison with the model where the intensities are scaled according to the OPR (red curve).
 In particular, it can be seen that the hyperfine satellites are in a regime of opacity where they still scale proportionally to the column density. The central component is however opaque and the two models differ by a factor $\sim$2. From this comparison, it can be inferred that if we consider the ratio of integrated intensities of the 86 and 110 GHz lines, this ratio
will only give a lower limit to the OPR. For the (0",0") position, this ratio is $\sim$2.9. Finally, if we only consider the most
optically thin hyperfine satellites, their integrated intensity ratio should be closer to the actual OPR. Considering these
components, we obtain an integrated intensity ratio 3.5$\pm$0.2. To conclude, the various ways of deriving the OPR
of NH$_2$D rule out a ratio lower than 3 and therefore favor the grain surface formation processes.

Considering these various estimates and since, to date,
the highest value expected for the OPR of NH$_2$D is 3, we use this latter value, in what follows, to derive the 
NH$_2$D total column density.
The total NH$_2$D column densities is thus log($N$) = $14.88^{+0.10}_{-0.14}$.

\subsubsection{NHD$_2$} 

\begin{figure}
\begin{center}
\includegraphics[angle=0,scale=0.35]{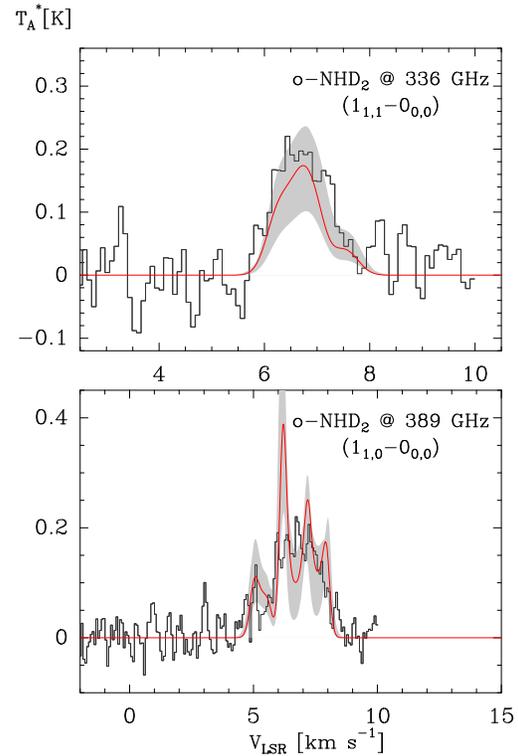}
\caption{Comparison between model (red curves) and observations (histograms) 
for the o--NHD$_2$ lines at 336 and 389 GHz observed towards B1b.}
\label{NHD2_center} \vspace{-0.1cm}
\end{center}
\end{figure}

Using the same methodology as for NH$_2$D, we re--analyzed the
o--NHD$_2$ (single position) observations published in
\citet{lis2006}. The spectroscopy of o--NHD$_2$ was obtained from the
\textit{splatalogue} database where the hyperfine structure
spectroscopy corresponds to the data published in
\citet{coudert2006,coudert2009}. As a first guess, we used the NH$_2$D
abundance profile discussed in the previous section, and tried to
scale it by a constant value in order to reproduce the two observed
lines at  336 and 389~GHz. By doing so, we could not find a reasonable
model. Indeed, a model that fits the  336 GHz line could be
obtained by scaling down the o--NH$_2$D abundance by a factor
4.5. However, this would lead to a model where the integrated area of
the 389  GHz line would be overestimated by a factor $\sim$3. As stated
by \citet{lis2006}, the 389  GHz line has a higher critical density than
the  336 GHz line. Hence, in order to simultaneously fit the two
o--NHD$_2$ lines, the NH$_2$D abundance profile has to be modified
with a global shift of the abundance towards lower densities,
i.e. larger radii. Such a model corresponds to the abundance profile
given in Fig. \ref{profils_B1} (right panel) and the corresponding fit
of the observations is shown in Fig. \ref{NHD2_center}.
As an alternative to this modification, 
we checked other possibilities that would enable to define
NH$_2$D and NHD$_2$ abundance profiles that would only differ by a scaling factor. We found that assuming 
a constant gas temperature around 10K would allow such a solution. However, given the 
constraints obtained on the dust temperature across the B1b region \citep[see][]{daniel2013} 
and given the high densities in the innermost part of the core, this solution was discarded.

The NHD$_2$ observations in B1b were already discussed in
\citet{lis2006}.  As mentioned earlier, in the latter study, the
column density estimated separately from the 336 or 389 GHz lines, using an LTE
approximation, were found to disagree by a significant factor. Indeed,
these beam averaged column densities were respectively found to be $1.6 \times 10^{14}$ and $2.9 \times
10^{13}$ cm$^{-2}$, i.e.log($N$) = $14.20$ and $13.46$.  
Here, the o--NHD$_2$ column density is estimated 
to log($N$) = $13.90^{+0.38}_{-0.26}$.
Finally, a previous analysis by \citet{lis2006} 
have shown that in B1b, the OPR is consistent with the ratio of 
the nuclear spin statistical weights, i.e. 2:1, within error bars.
With such a ratio, the total NHD$_2$ column densities is thus
log($N$) = $14.08^{+0.38}_{-0.26}$.

\subsubsection{ND$_3$}

To model the ND$_3$ transition, we scaled the o--NH$_2$D abundance
profile by an overall factor.  In fact, the exact shape of the
abundance profile is not constrained since the observed line is
optically thin and because observations are available only towards a
single position. As a consequence, alternate abundance profiles
would provide similarly good fits to the observations. Hence, the only
parameter which is well constrained by the model is the column
density.  The m--ND$_3$ column density is estimated to log($N$) =
$12.71^{+0.17}_{-0.20}$.
In a previous study, \citet{lis2002} reported a ND$_3$ column density
of $2\pm0.9 \times 10^{12}$ cm$^{-2}$, ie. log($N$) = $12.30^{+0.16}_{-0.16}$. This value
corresponds to the total ND$_3$ column density, i.e. which consider all the spin isomers, and
is averaged over the beam size. With the current model, the column density averaged over a beam of 25"
lead to a column density for m--ND$_3$ of  $2.7 \times 10^{12}$ cm$^{-2}$. Assuming, that the ortho, meta and para
spin isomers are populated according to their spin statistical weights, i.e. 16:10:1, we obtain a total column density of 
$7.3 \times 10^{12}$ cm$^{-2}$ which is a factor $\sim$3.5 higher than the estimate of \citet{lis2002}.
Note however that in \citet{lis2002}, the column density estimate assumes that the various spin isomers are 
thermalized at the gas temperature. At 5K and 10K, this implies that $\sim$60\% and 45\% of the molecules
are, respectively,  in the meta modification. In the current estimate, we assume that the proportion of molecules in the meta
modification is independent of the temperature and is set at 37\%. Hence, a factor $\sim$1.5 in the two estimates
comes from the hypothesis respectively made on the partition functions.

Finally, assuming that the ortho, para and
meta ND$_3$ states are populated according to the ratios 16:1:10, we derive
log($N$) = $13.14^{+0.17}_{-0.20}$.

\begin{figure}
\begin{center}
\includegraphics[angle=0,scale=0.35]{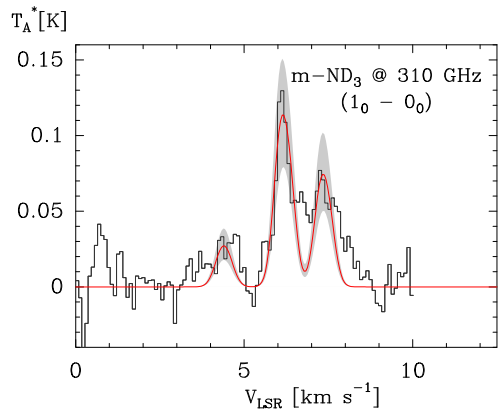}
\caption{Comparison between model (red curves) and observations (histograms) 
for the m--ND$_3$ line at 309.9 GHz observed towards B1b.}
\label{ND3_center} \vspace{-0.1cm}
\end{center}
\end{figure}

\subsubsection{Discussion}

 Deviations of the OPR from the statistical weights ratio are expected if the molecules form in the
gas-phase owing to nuclear--spin selection rules. Nuclear--spin effects
in the gas-phase have, however, a complex dependence on density,
temperature and time \citep{rist2013,faure2013,sipila2015}. Here, we have thus assumed
that the various nuclear--spin species of all four ammonia isotopologues are populated according to their
spin statistical weights, in order to translate the observed column densities
to total column densities. 
In the case of the main isotopologue of ammonia, the p--NH$_3$ column density was estimated to
be log($N$) = $14.74^{+0.25}_{-0.13}$ \citep{daniel2013}, which
translates to a total column density of log($N$) =
$15.04^{+0.25}_{-0.13}$ assuming a statistical OPR of
1:1. 

\begin{table*}
\caption{Ratio of column density of the various isotopologues of NH$_3$, obtained for the B1b and 16293E molecular clouds. The column
\textit{new} refers to the current results while the column \textit{old} corresponds to the values reported by \citet{roueff2005}.}
\begin{center}
\begin{tabular}{c|cc|cc|cc|cc|}
              & \multicolumn{2}{c}{[NH$_2$D]/[NH$_3$]} & \multicolumn{2}{c}{[ND$_2$H]/[NH$_2$D]} & \multicolumn{2}{c}{[ND$_3$]/[NHD$_2$]} & \multicolumn{2}{c}{[ND$_3$]/[NH$_3$]} \\
               & old & new & old & new & old & new & old & new   \\ \hline
B1b        & 0.23$\pm$0.05 & 0.69$^{+0.24}_{-0.19}$ & 0.15$\pm$0.03 & 0.16$^{+0.22}_{-0.03}$ & 0.033$\pm$0.01 & 0.11$^{+0.09}_{-0.04}$ & 1.1$\pm$0.5 $\times 10^{-3}$  &   1.3$\pm$0.6 $\times 10^{-2}$ \vspace{0.1cm} \\
16293E  & 0.19$\pm$0.05 &   ...   & 0.22$\pm$0.05 & 0.21$^{+0.08}_{-0.06}$ & 0.024$\pm$0.01 & 0.05$^{+0.02}_{-0.01}$ & 1.0$\pm$0.5 $\times 10^{-3}$  &  ... \\ 
\end{tabular}
\end{center}
\label{default}
\end{table*}%

The fractionation ratios derived from the current analysis are
reported in the first entry of Table~\ref{default}.
Note that with respect to the previous estimate 
reported by \citet{roueff2005} and 
displayed in this table, the current analysis leads to larger ratio, the updated values being
higher by factors in the range 1--10.
We thus obtain for Barnard 1b NH$_2$D/NH$_3\sim$0.69,
NHD$_2$/NH$_2$D$\sim$0.16 and ND$_3$/NHD$_2\sim$0.11. These ratios
provide very useful constraints for astrochemical models and they
should help, in particular, to disentangle between a gas-phase and a
grain surface formation of the ammonia isotopologues. The gas-phase
model of \cite{roueff2015} was shown to reproduce the observed column
density ratios of all four ammonia isotopologues in B1b and 16293E
within a factor of 3. The updated observed value for NH$_2$D/NH$_3$ would be
however underpredicted by a factor of $\sim 5$. On the other hand, it is interesting to note that a
purely statistical model\footnote{A purely statistical
  approach predicts the following ratios: NH$_2$D/NH$_3$=3$\times$D/H;
  NHD$_2$/NH$_2$D=D/H; ND$_3$/NHD$_2$=$\frac{1}{3}$$\times$D/H} of grain surface formation (i.e. through
statistical hydrogenation/deuteration of solid atomic nitrogen)
reproduces the observed ratios within the error bars, provided that the
accreting atomic D/H ratio is $\sim$0.2. Such a high
value for the D/H ratio is actually predicted in regions with high
density and heavy depletion, such as prestellar cores
\citep{roberts2003}. Clearly, more observations of the four ammonia
isotopologues, in several other sources, as well as other related
species such as ND and NHD, are necessary to make progress.

\subsection{16293E}

The IRAS 16293E source is a prestellar core located close to the IRAS 16293-2422 protostar.
Hence, its physical and chemical properties are influenced by this source, 
mainly because of the interaction with the ouflows that emanate from it \citep{castets2001,lis2002b}.

To model the emission of the ammonia isotopologues observed towards
the 16293E core, we adopt the source structure described in Bacmann et
al. (2015).  Based on ground--based and space--based observations of
the dust continuum emission at various wavelengths, the center of the
density profile is assumed to be at the position $\alpha_{2000} =
16^h32^m28.8^s$, $\delta_{2000} = -24^\circ29'04.0''$ \citep{bacmann2015}.
The reference position of the observations is 
$\alpha_{2000} =
16^h32^m29.47^s$, $\delta_{2000} = -24^\circ28'52.6''$ \citep{roueff2005,lis2006}.

In order to constrain the behaviour of the abundance, we start by
considering the o--NHD$_2$ isotopologue, for which we have maps of the
 336 GHz line emission obtained with APEX \citep{gerin2006}, as well as
CSO observations of the  336 and 389 GHz lines obtained close to the
dust emission peak \citep{lis2006}. As for the B1b modeling, it is
confirmed that these two lines are good probes of the radial
variations of the abundance profile, because of their
respective critical densities and because of the difference of opacity
of the lines.

\subsubsection{NHD$_2$}

\begin{figure}
\begin{center}
\includegraphics[angle=0,scale=0.35]{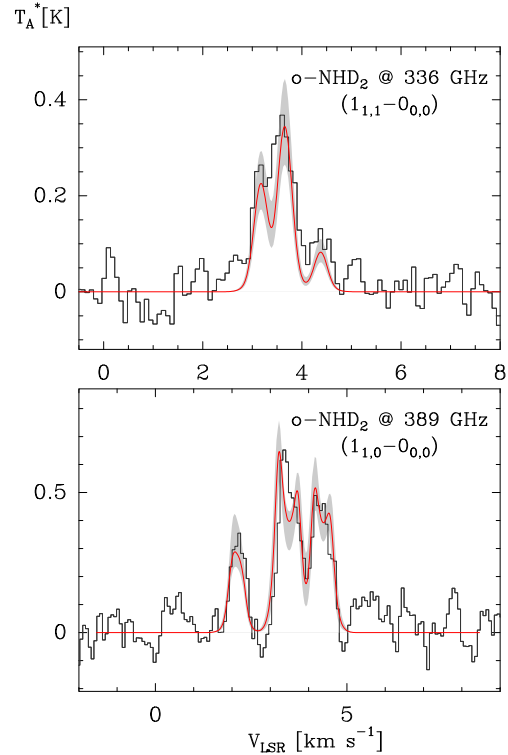}
\caption{Comparison between model (red curves) and observations (histograms) 
for the o--NHD$_2$ line at 336 GHz and 389 GHz observed with the CSO towards 16293E.
The position observed is at an offset of (-10",-10") with respect to the reference position 
$\alpha_{2000} = 16^h32^m29^s.47$, 
$\delta_{2000} = -24^\circ28'52.6''$.
}
\label{L1689N_oNHD2_center} \vspace{-0.1cm}
\end{center}
\end{figure}

\begin{figure}
\begin{center}
\includegraphics[angle=0,scale=0.7]{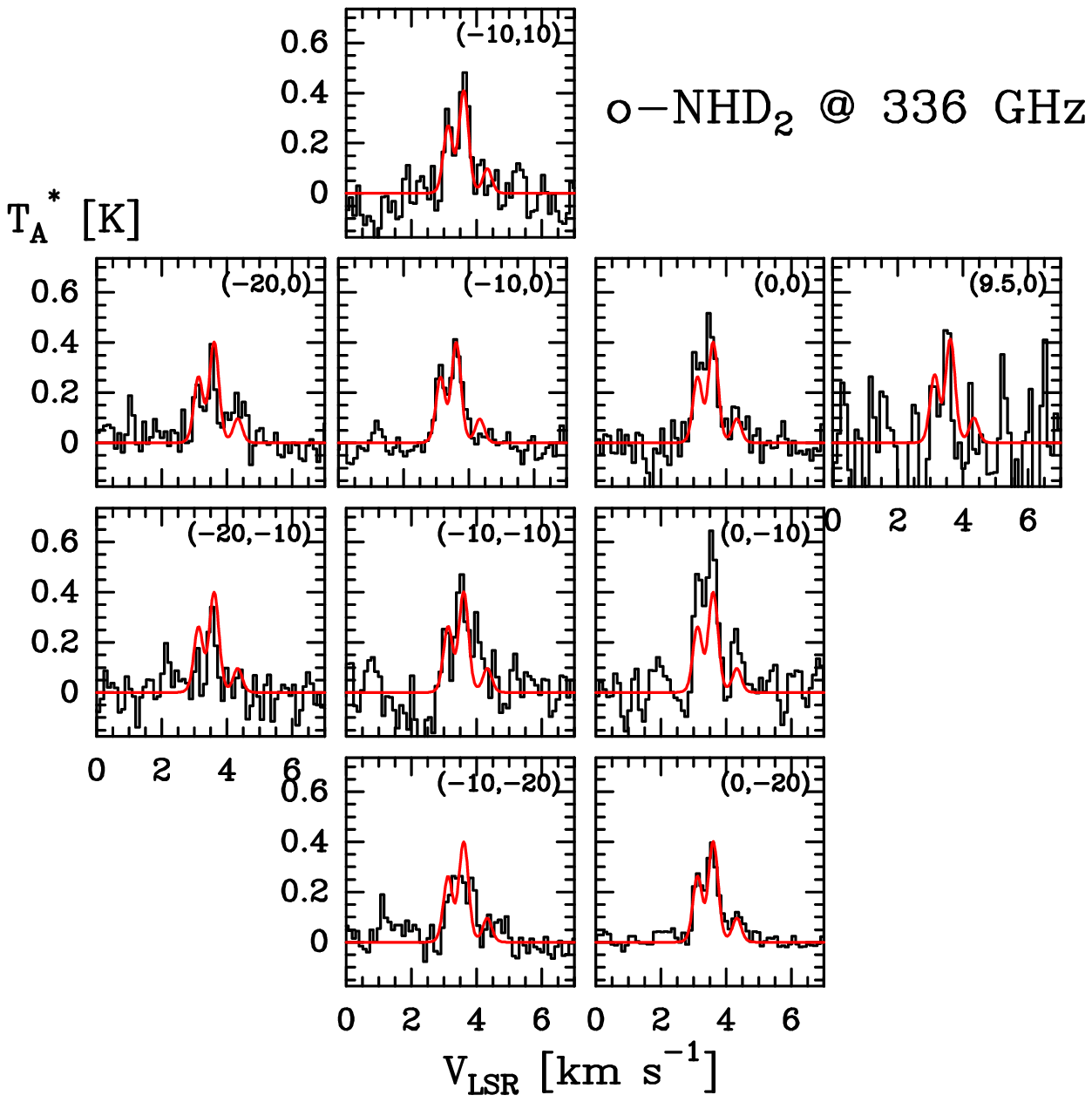}
\caption{ Comparison between model (red curves) and observations (histograms) 
for the o--NHD$_2$ line at 336 GHz observed with APEX towards 16293E. In each panel,
the offset is indicated according to the reference position $\alpha_{2000} = 16^h32^m29^s.47$, 
$\delta_{2000} = -24^\circ28'52.6''$}
\label{L1689N_oNHD2_map} \vspace{-0.1cm}
\end{center}
\end{figure}

\begin{figure}
\begin{center}
  \includegraphics[angle=0,scale=0.3]{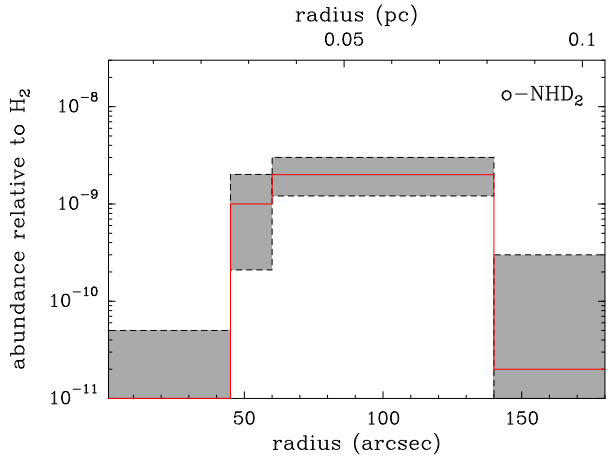}
\end{center}
\caption{
  Abundance profile of o--NHD$_2$ (red curve) as a function of
  radius in the prestellar core 16293E. The grey areas show the
  confidence zone for the abundance at every radius.}
\label{profil_16293E}
\end{figure}

The comparison between the model and observations for the 336 and 389
GHz lines is given in Fig. \ref{L1689N_oNHD2_center}. The map of the
336 GHz line obtained with APEX is reported in
Fig. \ref{L1689N_oNHD2_map}. 
The abundance profile, which corresponds to these models, is given in
Fig.  \ref{profil_16293E}.

\begin{figure}
\begin{center}
\includegraphics[angle=0,scale=0.7]{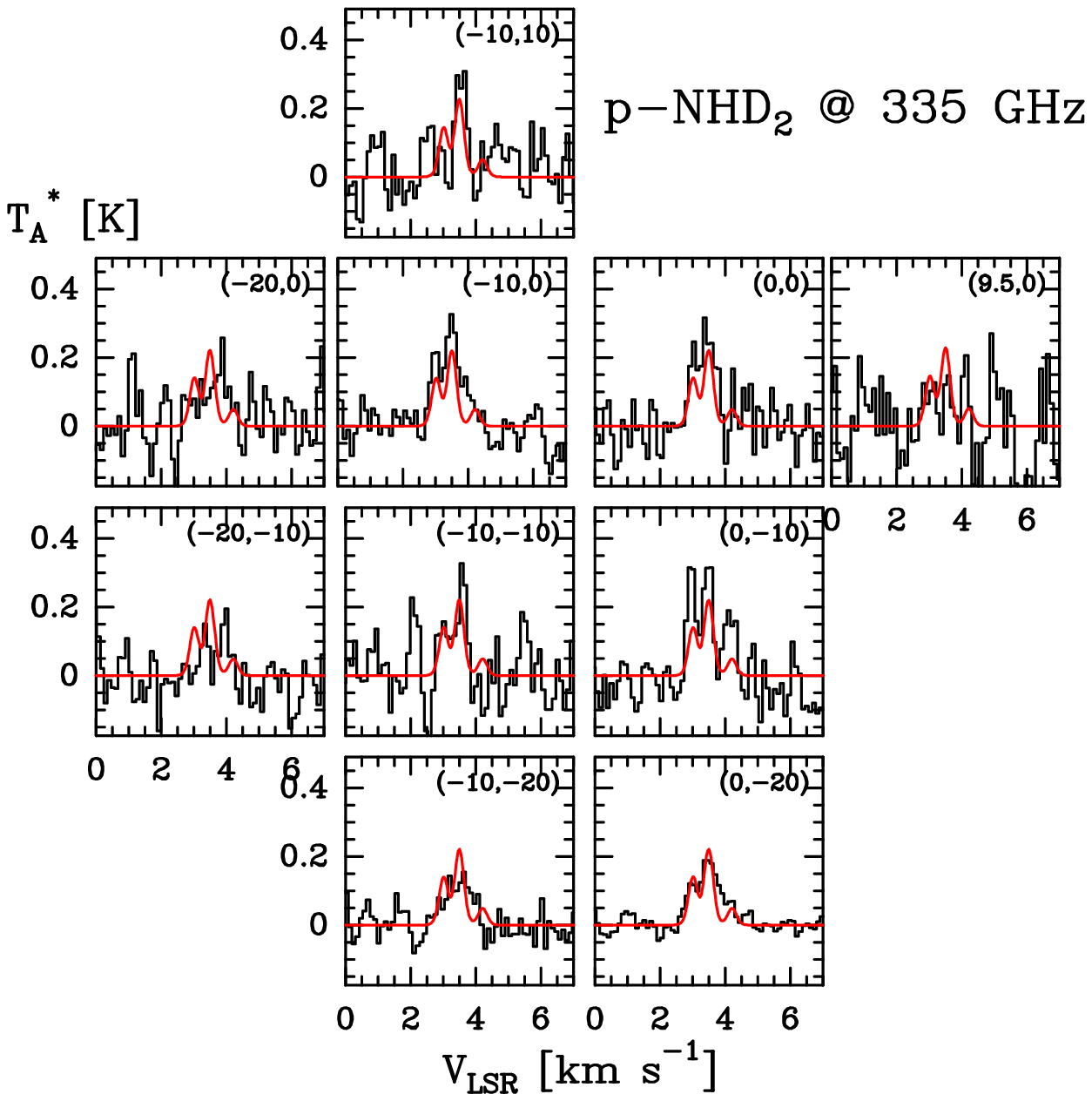}
\caption{Comparison between model (red curves) and observations (histograms) 
for the p--NHD$_2$ line at 335 GHz observed with APEX towards 16293E. In each panel,
the offset is indicated according to the reference position $\alpha_{2000} = 16^h32^m29^s.47$, 
$\delta_{2000} = -24^\circ28'52.6''$}
\label{L1689N_pNHD2_map} \vspace{-0.1cm}
\end{center}
\end{figure}

As outlined in \citet{gerin2006}, the
o--NHD$_2$ and the dust emission peak are offset by $\sim$10''. With
the current spherical model, such an asymmetry cannot be taken into
account and the model shown in Fig. \ref{L1689N_oNHD2_map} thus only
reproduces qualitatively the overall behaviour of the o--NHD$_2$
emission.  Finally, assuming an OPR of 2:1, i.e. given
by the nuclear spin statistical weights, we obtain a good fit to the
APEX p--NHD$_2$ map. The comparison between model and observations is
reported in Fig. \ref{L1689N_pNHD2_map}. The OPR is
similar to the one obtained by \citet{gerin2006} where the abundance
ratio was derived from the line intensity ratio, which is accurate in
the case of optically thin lines.

The o--NHD$_2$ column density is log($N$) = $13.82^{+0.14}_{-0.15}$ and with a 2:1 ratio
between the ortho and para species, this gives a total column density of log($N$) = $13.99^{+0.14}_{-0.15}$.

\subsubsection{NH$_2$D}

\begin{figure}
\begin{center}
\includegraphics[angle=0,scale=0.40]{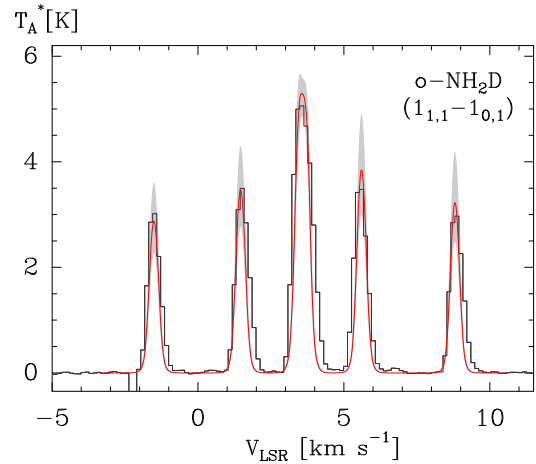}
\caption{Comparison between model (red curves) and observations (histograms) 
for the o--NH$_2$D line observed towards 16293E at the reference position.}
\label{L1689N_oNH2D_center} \vspace{-0.1cm}
\end{center}
\end{figure}

The o--NH$_2$D ($1_{1,1}$--$1_{0,1}$) line was observed at the single position 
$\alpha_{2000} = 16^h32^m29^s.47$, $\delta_{2000} = -24^\circ28'52.6''$,
which is offset by $\sim$15" from the center of our model.
This o--NH$_2$D observation can be satisfactorily reproduced by applying an overall scaling factor to the o--NHD$_2$ 
abundance profile discussed in the previous section. The comparison between the model and observations
is shown in Fig. \ref{L1689N_oNH2D_center}. 
The o--NH$_2$D column density is log($N$) = $14.54^{+0.13}_{-0.08}$ and with a 3:1 ratio
between the ortho and para species, this gives a total column density of log($N$) = $14.66^{+0.13}_{-0.08}$.

\subsubsection{ND$_3$}

\begin{figure}
\begin{center}
\includegraphics[angle=0,scale=0.70]{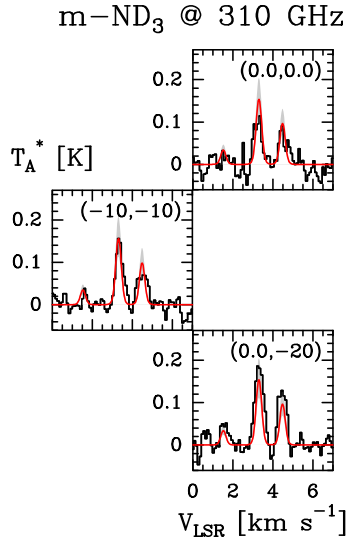}
\caption{Comparison between model (red curves) and observations (histograms) 
for the m--ND$_3$ line observed towards 16293E.
The offsets are indicated according to the reference position $\alpha_{2000} = 16^h32^m29^s.47$,
$\delta_{2000} = -24^\circ28'52.6''$.
}
\label{L1689N_mND3_map} \vspace{-0.1cm}
\end{center}
\end{figure}

As for NH$_2$D, we model the ND$_3$ transition by applying an overall scaling factor to the o--NHD$_2$ abundance
profile. The m--ND$_3$ column density is log($N$) = $12.27^{+0.15}_{-0.16}$ and
with a 16:10:1 ratio between the ortho, meta and para species, this
gives a total column density of log($N$) = $12.70^{+0.15}_{-0.16}$.
The comparison between the model and observations
is shown in Fig. \ref{L1689N_mND3_map}.

\subsubsection{Discussion}

The column densities derived for the four ammonia isotopologues in
16293E are reported in the second entry of Table~\ref{default}.
Again, our new analysis does not change much the fractionation
ratios with respect to the study by \cite{roueff2005}. We thus obtain
for 16293E the ratios NHD$_2$/NH$_2$D$\sim$0.21 and
ND$_3$/NHD$_2\sim$0.05, which are in fact very similar to those
derived in B1b. Such small variations in the
fractionation ratios suggest similar physical conditions in both
sources and, therefore, similar chemical pathways. We note that these
fractionation ratios can be reproduced within a factor of 3 by the
gas-phase model of \cite{roueff2005}, but also by a purely statistical
grain surface model. Once again, more observations are required to
make further progress in our understanding of the deuterium
fractionation of ammonia and to rule out one of the two possible
scenario.

\section{Conclusions} \label{conclusion}

We used the state--of--the--art potential energy surface that
describes the NH$_3$--H$_2$ interaction \citep{maret2009} in order to
determine collisional rate coefficients for the ND$_2$H and ND$_3$
isotopologues.  We subsequently used the Close--Coupling method to
calculate a set of collisional rate coefficients for ND$_2$H, which
applies to both the para and ortho symmetries of this molecule. For
ND$_3$, we made specific calculations for the ortho and para
species. A few test calculations showed that this latter set was also
suitable to treat the meta symmetry of ND$_3$.

The ND$_2$H--H$_2$ rate coefficients were compared to earlier
calculations based on the same PES but using a reduced basis for H$_2$
\citep{wiesenfeld2011}. We found that the new set shows differences of up
to a factor $\sim$2.5 with respect to the old data. Moreover, the
rates which are the most affected are those of highest magnitude.  In
the case of ND$_3$, we compared our calculations with ND$_3$--He
calculations \citep{machin2006}.  As previously found for ND$_2$H
\citep{wiesenfeld2011} or NH$_2$D \citep{daniel2014}, we find large
differences between the H$_2$ and He rate coefficients with variations of up to two
orders of magnitude.

Finally, we used these new rate coefficients, as well as the 
NH$_2$D--H$_2$ calculations published earlier \citep{daniel2014}, in order to
re--interpret the observations available in the literature of the
various deuterated isotopologues of NH$_3$
\citep{roueff2000,lis2002,roueff2005,gerin2006,lis2006}. We focused on
the B1b and 16293E prestellar cores and we used a comprehensive
radiative transfer analysis based on the latest estimates of the
physical structure of these objects \citep{daniel2013,bacmann2015}. By
comparison with the earlier estimates of the column densities, based
on approximate methods which did not solve the coupled set of
radiative transfer and statistical equilibrium equations, we obtained
a modest revision of the column densities. Additionally, the column
density ratios of the various deuterated isotopologues agree, within a
factor of $\sim$2--3, with the previous values published for these two
objects. This shows that the LTE method, which is often used to carry
out a first order analysis independent of collisional rates, is a
reliable alternative to analyze the emission of the NH$_3$ deuterated
isotopologues, at least if one is concerned with obtaining the order
of magnitude of the deuterium enrichment. However, in the case a
factor of 2 accuracy is required, as when constraints are to be set on
the abundance ratios of molecular spin isomers, the new rate
coefficients have to be used in a model where the molecular
excitation is solved.

\section*{Acknowledgments}

All (or most of) the computations presented in this paper were
performed using the CIMENT infrastructure
(https://ciment.ujf-grenoble.fr), which is supported by the
Rh\^one-Alpes region (GRANT CPER07\_13 CIRA: http://www.ci-ra.org).
D.L. support for this work was provided by NASA (\emph{Herschel} OT
funding) through an award issued by JPL/Caltech.  This work has been
supported by the Agence Nationale de la Recherche (ANR-HYDRIDES),
contract ANR-12-BS05-0011-01 and by the CNRS national program
``Physico-Chimie du Milieu Interstellaire''. The authors thanks Q. Ma, P. Dadgigian
and A. Van der Avoird for providing some ND$_3$ cross sections. We also thank J. Harju for 
constructive discussions.

\bibliographystyle{mn2e}
\bibliography{biblio}

\appendix

\section{ND$_3$ rotational levels }

ND$_3$ rotational levels can be inferred from ammonia spectroscopy
\citep{Townes}.  In their ground electronic state, both NH$_3$ and
ND$_3$ are symmetric top with a C$_{3v}$ pyramidal symmetry. The
symmetric top rotational levels are labeled $J_K$ where $J$ stands for
the rotational momentum quantum number and $K$ is its absolute
projection along the symmetry axis.  The rotation states are
eigenstates of the symmetric top Hamiltonian, namely $\ket{J K m}$ where
$m$ is the projection quantum number of the angular momentum along any
space-fixed axis \citep{Green76}.  All rotational levels $J_K$ are
degenerate in $m$.

As for ammonia, the nitrogen atom can tunnel back and forth through the deuterium plane. In this umbrella vibrational mode ($\nu_2$)  the potential presents a double minimum along the inversion coordinate. This results in the splitting of all rotational levels  into two inversion levels (states) respectively symmetric $\ket{+}$ and antisymmetric $\ket{-}$ with respect to the inversion coordinate. Although we neglect any vibrational coupling during the collision, the weak energy splitting between inversion levels is set to  a constant $0.053$ cm$^{-1}$ \citep{Tkac2014,Fusina1986397,Fusina1994464} independantly of the rotational level. The lower inversion level $\ket{+}$ is symmetric whereas the upper level  $\ket{-}$ is antisymmetric with respect to the inversion coordinate.  
 
\subsection{Deuterium permutation symmetry}
Because D nuclei have a nuclear spin I = 1, they obey Bose--Einstein statistics and the total internal wave-function of ND$_3$ is unchanged by permutation of two identical D. Thus the determination of all allowed internal states requires the description of the deuterium  nuclear-spin states together with the symmetry under permutation of the three identical D atoms for both motional and nuclear-spin states. From Bose-Einstein statistics  one derives new statistical weights of the rotation-inversion levels  compared to NH$_3$ \citep{Bunker,Hugo2009}.

In the ground electronic state, the molecular symmetry group of ND$_3$ is C$_{3v}$ and the complete nuclear spin permutation group is S$_3$ \citep{Bunker}. The associated irreducible representations correspond to A$_1$, A$_2$ and E symmetry species. The one dimension symmetry adapted basis of A$_1$ and A$_2$ symmetry are wave-functions respectively symmetric and antisymmetric under permutation of two nuclei whereas in any two dimension basis that spans the E representation, the permutation matrix is non diagonal. In the following we recall the symmetry species of each rotational, inversion and nuclear-spin state.

%In the one dimension A$_1$ irreducible representation any interchange of two deuterons leaves the wave-function unchanged whereas in the A$_2$ representation it changes sign. In the two dimension E representations the interchange  of two deuterons ....

\subsection{Permutation symmetry adapted motional states}
As for ammonia, the symmetric top rotational states transform as follows under permutation $\cal{P}$ of two deuterons \citep{Townes,Bunker}:
\[ {\cal{P} }\ket{JKm} = (-1)^{J}e^{-2in\pi K/3} \ket{J-Km} ; n=1,2,3 \]
Thus symmetry adapted rotational states are combinations of $\ket{JKm}$ and $\ket{J-Km}$ states \citep{Townes}.  Symmetric and antisymmetric combinations of $\ket{JKm}$ and $\ket{J-Km}$, for which $K$ is a multiple of $3$, span the A$_1$ and A$_2$ representations :
\[ \ket{JKm\alpha} =\sqrt{1 \over 2 (1 + \delta_{K0})} (\ket{JKm} + \alpha \ket{J-Km}) ; \alpha=\pm 1 \] 
\[ {\cal{P}}  \ket{JKm\alpha}  = \alpha (-1)^J \ket{JKm\alpha} \]

For each rotational level $J_K$ the permutation symmetry  species A$_1$ or  A$_2$ are given by the sign of $\alpha (-1)^J $. Rotational states of both symmetries exist with the exception of $K=0$   for which  $\alpha = 1$ and  even $J$ states span the A$_1$ symmetry whereas odd $J$ states span the A$_2$ symmetry.    
Any basis ($\ket{JKm}$ ; $\ket{J-Km}$) of rotational states for which $K$ is not a multiple of $3$ spans a representation of E symmetry. 

For the inversion motion the symmetric and antisymmetric vibrational states $\ket{+}$ and $\ket{-}$ span  respectively the A$_1$ and $A_2$ symmetry \citep{Townes}:
\[ {\cal{P}} \ket{\pm} = \pm \ket{\pm}\]

\subsection{Symmetry adapted nuclear-spin states}
The total nuclear-spin wave-function is obtained from the coupling of three deuterium nuclear-spin momentum with nuclear-spin quantum number $I=1$. The direct product of the three deuteron nuclear-spin states (with $m_I= -1, 0, 1$) where $m_I$ is the nuclear-spin projection quantum number, gives  $3^3=27$ deuterium spin states for ND$_3$. In terms of nuclear-spin angular momentum, from the single deuteron  rotation group irreducible representation, D$_1$ one builds  the complete nuclear-spin representation as the direct product: $(D_1)^3$ which decomposes as follows on the total nuclear-spin irreducible representations D$_I$ of total quantum number I \citep{Atkins}: 
\[ D_1^3 = D_3 + 2 D_2 + 3 D_1 + D_0 \]
Each of the angular momentum basis characterized by $(I,m_I)$ spans
the permutation symmetry species $\Gamma_{ns,I}$ as follows
\citep{Hugo2009,machin2006}: $I=3 \rightarrow A_1$, $I=2 \rightarrow
E$, $I=1 \rightarrow A_1+E$ and $I=0 \rightarrow A_2$.  Thus the
angular momentum representation characterized by the total
nuclear-spin $I$ generate the permutation symmetry representations
$\Gamma_{ns,I}$ \citep{Hugo2009}:
\begin{eqnarray}
 D_3 & = & 7 A_1 \\
 2 D_2 & = & 5 E \\
 3 D_1 & = & 3 A_1 + 3E \\
 D_0 & = & A_2
\end{eqnarray}
In terms of permutation group, the total  nuclear-spin representation  $\Gamma_{tot}$ decomposes on the irreducible representations $\Gamma_{ns}=A_1, E, A_2$ as follows: 
 \[\Gamma_{tot} = {\sum}_{I,\Gamma_{ns,I}}(2I+1)\Gamma_{ns,I}=10A_1 + 8E+ A_2\]
 For the total nuclear spin basis (of 27 nuclear-spin states), the
 symmetry degeneracies are respectively $(g_{A_1}=10,
 g_{E}=16,g_{A_2}=1)$. We note that the E representation is of
 dimension 2 and therefore the number of nuclear-spin basis states of
 E symmetry is 16.  From this decomposition we identify the three
 ND$_3$ nuclear-spin modifications as defined by \cite{Maue}, in
 decreasing order of the symmetry degeneracy $g_{ns}$: ortho
 (E, $g_{ns}=16$), meta (A$_1$, $g_{ns}=10$) and para
 (A$_2$, $g_{ns}=1$).

\subsection{Nuclear-spin statistics}
The total nuclear-spin rotation inversion wave-functions are obtained from the direct product of nuclear-spin states with the rotation states and vibration state. The direct product of two S$_3$ irreducible representations generates a new representation shown in table \ref{table:directproduct} \citep{Bunker,Atkins}. 
\begin{table}
\caption{Direct product representation of S$_3$}
\begin{center}
\begin{tabular}{cccc}
\hline
$\Gamma \otimes \Gamma'$ & A$_1$ & A$_2$  & E \\ \hline
A$_1$           &  A$_1$ & A$_2$ & E \\
A$_2$           &  A$_2$ & A$_1$ & E\\
E                   &  E         & E         &  A$_1$+A$_2$+E \\
\end{tabular}
\end{center}
\label{table:directproduct}
\end{table}%
Because of Bose-Einstein principle the complete wave-function is
unchanged by the interchange of two deuterium. In terms of permutation
symmetry it belongs to the A$_1$ symmetry of S$_3$. Therefore given
the symmetry species A$_1$ and A$_2$ of ND$_3$ inversion states $\ket{+}$
and $\ket{-}$ , each inversion state can only combine to
nuclear-spin-rotation states of the same symmetry. The symmetry
species of the nuclear-spin-rotation states can also be deduced from
the direct product of Table \ref{table:directproduct}. The para and
meta nuclear-spin state of symmetry A$_2$ or A$_1$ can combine to any
A symmetry rotational state with $K=3n$, whereas the ortho
nuclear-spin states of symmetry E can combine to any E symmetry
rotational state with $K \neq 3n$. The symmetry of the resultant
direct product nuclear-spin-rotation state is either A$_1$ or A$_2$.
 
In reference to previous description of ammonia
\citep{Green76,Offer89,Rist93,ma2015}, for each modification, the
symmetry adapted ND$_3$ nuclear-spin-rotation-inversion states are
commonly labeled by $\ket{J K m \epsilon } \ket{\pm}$ where $\epsilon = \pm 1$
specifies the symmetry of the nuclear-spin-rotation state: ${\cal{P}}
\ket{JKm\epsilon} = \epsilon (-1)^J \ket{JKm\epsilon} $ for the meta and ortho
modifications and ${\cal{P}} \ket{JKm\epsilon} = - \epsilon (-1)^J \ket{JKm\epsilon} $
for the para modification. Therefore each nuclear-spin
rotation inversion state is uniquely defined by the label 
$\ket{JKm\epsilon} \ket{\pm}$. 
Conversely, for a given spin isomer, it is only necessary to specify either $\epsilon$
or the inversion state since these quantities are related through the relations : 
$\epsilon = -(-1)^J \ket{\pm}$ for p--ND$_3$ and 
$\epsilon = (-1)^J \ket{\pm}$ for o--ND$_3$ or m--ND$_3$.
We note that this definition of $\epsilon$ is
consistent with the previous description of NH$_3$ spin-rotation
states but leads to an opposite inversion symmetry for the meta and
ortho modifications \citep{ma2015}.

For rotational states with $K=0$, as the spin-rotation states $\ket{J0m\epsilon}$ are restricted to $\epsilon = 1$ only one of the inversion state $\ket{\pm}$  fulfills Bose-Einstein principle; 
$\pm (-1)^J = 1$  for the A$_1$ meta spin states and $\pm (-1)^J = -1$ for the A$_2$ para spin states.  For all other rotational states with $K \neq 0$, both inversion states can combine to the rotation-nuclear-spin states.  A summary of the permutation symmetry of all nuclear-spin-rotation-inversion states with their nuclear-spin weights is given in table \ref{table:spin-rotation-inversion}. The spin statistics of all rotation-inversion states is deduced in table \ref{table:spin-statistic}. The consequent rotational energy level diagram is given by \cite{Tkac2014}.
\begin{table*}
\caption{Permutation symmetry of ND$_3$ nuclear-spin-rotation-inversion states: $ \ket{JKm\epsilon} \ket{\pm}$}
\begin{center}
\begin{tabular}{ccccccccc}
\hline
nuclear spin ($\Gamma_{ns}$) &$\Gamma_{rot}$ &   $\Gamma_{ns} \otimes \Gamma_{rot}$ &  $\Gamma_{inv} $  & $K$      & $J_{K=0}$&    $\epsilon$ &  $\ket{\pm}$   & g$_{ns}$ \\ \hline
Para (A$_2$)                                    & A$_2$                & A$_1$                                                       & A$_1$                     & $3n$         & $J$  odd  & $-(-1)^J$ &  $\ket{+}$    &        1     \\
                                                         & A$_1$                & A$_2$                                                       & A$_2$                      & $3n$         & $J$ even &$(-1)^J$  &   $\ket{-}$   &        1 \\ \hline

Meta (A$_1$)                                   &  A$_1$               & A$_1$                                                       & A$_1$                      & $3n$       & $J$ even  &$(-1)^J$ &   $\ket{+}$   &       10  \\
                                                         &  A$_2$               & A$_2$                                                       & A$_2$                      & $3n$       & $J$ odd   & $-(-1)^J$&   $\ket{-}$   &       10  \\ \hline

Ortho  (E)                                         & E                        & A$_1$                                                       & A$_1$                      & $3n\pm 1$  &  -       &  $(-1)^J$&   $\ket{+}$  &       8  \\
                                                         & E                        & A$_2$                                                       & A$_2$                      & $3n\pm 1$  &  -       &  $-(-1)^J$&  $\ket{-}$  &       8  \\
 \\
\end{tabular}
\end{center}
\label{table:spin-rotation-inversion}
\end{table*}%
\begin{table}
\caption{Nuclear-spin statistics of rotation-inversion states for ND$_3$ }
\begin{center}
\begin{tabular}{ccccccccc}
\hline
  $K$ & $J$   & $\epsilon$ & $\ket{\pm}$  & spin modification   & g$_{ns}$  \\ \hline
  $0$ & even & $1$  & $\ket{+}$ &  Meta                    & $10$            \\  
         & even & $1$ & $ \ket{-}$ &  Para                   & $1$            \\
         & odd  & $1$  & $\ket{+}$ &  Para                     & $1$            \\
         & odd  & $1$   & $\ket{-}$ &  Meta                    & $10$            \\ 
 $3n \neq 0$ &  all & $(-1)^J$  & $\ket{+}$          &      Meta                     & $10$            \\
                     &  all & $-(-1)^J$  & $\ket{+}$          &      Para                     & $1$            \\
                     &  all & $-(-1)^J$ & $\ket{-}$           &      Meta                     & $10$            \\
                     &  all & $(-1)^J$  & $\ket{-}$           &       Para                     & $1$            \\
  $3n \pm 1 $ &  all  & $(-1)^J$& $\ket{+}$ &  Ortho                     & $8$            \\
                      &  all  & $-(-1)^J$& $\ket{-}$  &  Ortho                     & $8$            \\

\end{tabular}
\end{center}
\label{table:spin-statistic}
\end{table}%

\label{lastpage}

\end{document}